\documentclass[preprints,article,submit,pdftex,moreauthors]{Definitions/mdpi}
\usepackage{cite}
\usepackage{algorithmic}
\usepackage{graphicx}
\usepackage{textcomp}
\usepackage{subfig}
\usepackage{amsmath}    
\usepackage{amsfonts}   
\usepackage{amssymb}    
\usepackage{mathtools}
\usepackage{physics}    
\firstpage{1} 
\pubvolume{1}
\issuenum{1}
\articlenumber{0}
\pubyear{2024}
\copyrightyear{2024}
\datereceived{ } 
\daterevised{ } 
\dateaccepted{ } 
\datepublished{ } 
\hreflink{https://doi.org/} 

\Title{Deep convolutional framelets for dose reconstruction in BNCT with Compton camera detector}

\TitleCitation{Deep convolutional framelets for dose reconstruction in BNCT with Compton camera detector}

\Author{Angelo Didonna$^{1}$*, Dayron Ramos Lopez$^{1,2}$*, Giuseppe Iaselli$^{1,2}$, Nicola Amoroso$^{1,3}$, Nicola Ferrara$^{1,2}$ and Gabriella Maria Incoronata Pugliese$^{1,2}$}

\AuthorNames{Angelo Didonna, Dayron Ramos Lopez, Giuseppe Iaselli, Nicola Amoroso, Nicola Ferrara and Gabriella Maria Incoronata Pugliese}

\AuthorCitation{
Didonna, A.; Lopez, D.R.; Iaselli, G.; Amoroso, N.; Ferrara, N.; Pugliese, G.M.I.}

\address{%
$^{1}$ \quad Istituto Nazionale di Fisica Nucleare, Sezione di Bari, 70125 Bari, Italy\\
$^{2}$ \quad Dipartimento Interateneo di Fisica, Università degli Studi di Bari Aldo Moro, 70125 Bari, Italy\\
$^{3}$ \quad Dipartimento di Farmacia-Scienze del Farmaco, Università degli Studi di Bari Aldo Moro, 70125 Bari, Italy\\
}

\corres{Author to whom correspondence should be addressed.}

\abstract{Boron Neutron Capture Therapy (BNCT) is an innovative binary form of radiation therapy with high selectivity towards cancer tissue based on the neutron capture reaction \textsuperscript{10}B(n,$\alpha$)\textsuperscript{7}Li, consisting in the exposition of patients to neutron beams after administration of a boron compound with preferential accumulation in cancer cells. The high linear energy transfer products of the ensuing reaction deposit their energy at cell level, sparing normal tissue.
    \newline
Although progress in accelerator-based BNCT has led to renewed interest in this cancer treatment modality, in vivo dose monitoring during treatment still remains not feasible and several approaches are under investigation.
    While Compton imaging presents various advantages over other imaging methods, it typically requires long reconstruction times, comparable with BNCT treatment duration.
    \newline
    This study aims to develop deep neural network models to estimate the dose distribution by using a simulated dataset of BNCT Compton camera images. The models pursue the avoidance of the iteration time associated with the maximum-likelihood expectation-maximization algorithm (MLEM), enabling a prompt dose reconstruction during the treatment.
    \newline
    The U-Net architecture and two variants based on the deep convolutional framelets framework have been used for noise and artifacts reduction in few-iterations reconstructed images, leading to promising results in terms of reconstruction accuracy and processing time.
}


\begin{document}



\section{Introduction}

\subsection{Boron neutron capture therapy}
Boron neutron capture therapy (BNCT) is a binary tumor-selective form of radiation therapy based on the very high affinity of the nuclide boron-10 for neutron capture, resulting in the prompt compound nuclear reaction \textsuperscript{10}B(n,$\alpha$)\textsuperscript{7}Li, and on the preferential accumulation of boron-containing pharmaceuticals in cancer cells rather than normal cells.\par
\begin{figure}
    \centering
    \includegraphics[width=0.9\linewidth, keepaspectratio]{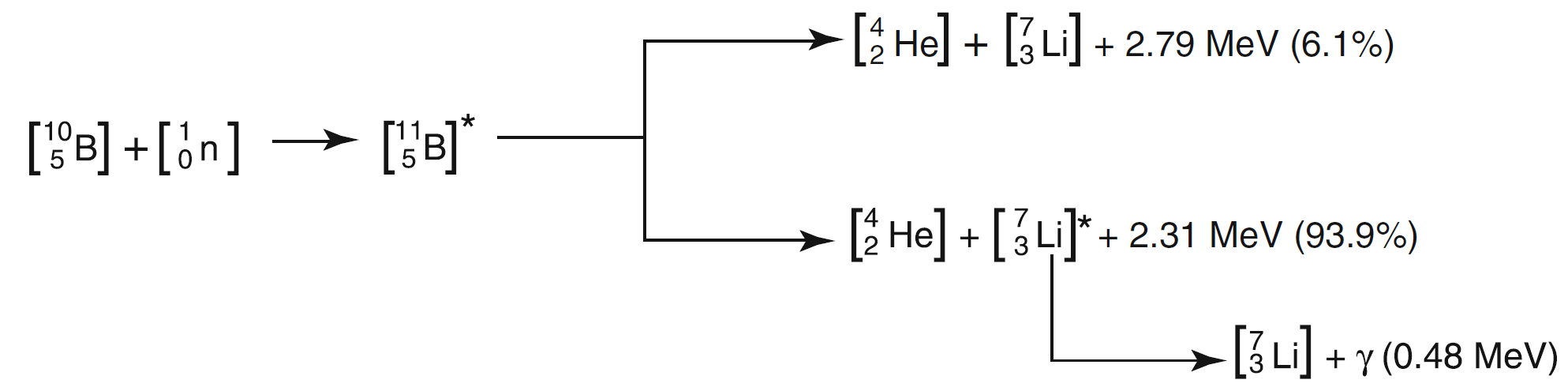}
    \caption[]{Boron neutron capture reaction \small{\citep{neutron_capture_springer}.}}
    \label{fig:bnct_reaction}
\end{figure}

{}
Neutron capture therapy (NCT) was first proposed  shortly after the discovery of the neutron by Chadwick in 1932 and the description of boron neutron capture reaction by Taylor and Goldhaber in 1935 \citep{IAEA_BNCT_2001}.
The reaction is illustrated in Figure~\ref{fig:bnct_reaction}: a \textsuperscript{10}B nucleus absorbs a slow or \emph{thermal} neutron (with energy $<0.5$ $eV$), forming for a brief time an highly excited \textsuperscript{11}B compound nucleus \citep{modern_nuclear_physics,handbook_nuclear_physics} which subsequently decays, predominantly disintegrating into an energetic alpha particle back to back with a recoiling \textsuperscript{7}Li ion 
, with a range in tissue of $\approx{}9$ $\mu{}m$ and $\approx{}5$ $\mu{}m$, respectively, and  a high linear energy transfer (LET) of $150$ $keV/\mu{}m$ and  $175$ $keV/\mu{}m$, respectively. The $Q$ value of the reaction is about $2.79$ $MeV$ \citep{podgorsak_radiation_physics_medical}. In about $94\%$ of the cases, the litium nucleus is in an excited state and immediately emits a $0.478$ $MeV$ gamma photon, leaving a combined average kinetic energy of about $2.31$ $MeV$ to the two nuclei.
\par{}
The selectivity of this kind of therapy towards cancer tissue is a consequence of the small combined range of \textsuperscript{7}Li and the alpha particle of $\approx{}12-14$ $\mu{}m$, which is comparable with cellular dimensions. By concentrating a sufficient number of boron atoms within tumor cells, the exposure to thermal neutrons may result into the their death with low normal tissue complications, because of the higher radiation dose imparted to cancer cells relative to adjacent normal cells.
After the administration of the boron compound to the patient, a proton or deuteron beam emitted from an accelerator is focused on a neutron emitting target. A suitable neutron beam is obtained with a beam shaping assembly (BSA) and is directed into a treatment room in which a patient is precisely placed. Typically the full dose is administered in one or two application of $30-90$ min in BNCT \citep{IAEA_BNCT_2023}, while in traditional photon treatments the dose is generally split into multiple fractions administered over a period of $3-7$ weeks. Therapeutic efficacy requires an absolute boron concentration higher than $20$ ppm and high \emph{tumor to normal tissue} ($T/N$) and \emph{tumor to blood} ($T/B$) \emph{concentration ratios}\footnote{They are also denoted by $C_{T}/C_{N}$ and $C_{T}/C_{B}$, where $C_{T}$, $C_{N}$ and $C_{B}$ are the boron concentrations in tumor tissue, normal tissue and blood, respectively.} (ideally $T/N>3$ and necessarily $T/N>1$ \citep{IAEA_BNCT_2023}).
\par{}
The dosimetry for BNCT is much more complex than for conventional photon radiation therapy. While in standard radiotherapy X-rays mainly produce electrons that release all their kinetic energy by ionization, in BNCT there are four main different radiation components contributing to the absorbed dose in tissue \citep{IAEA_BNCT_2023}. The \emph{boron dose} $D_{B}$, which is the therapeutic dose, is related to the locally deposited energy of about $2.33$ $MeV$ by the emitted alpha particle and the recoiling \textsuperscript{7}Li ion in the boron neutron capture reaction \textsuperscript{10}B(n,$\alpha$)\textsuperscript{7}Li.
\par{}
It can be shown \citep{IAEA_BNCT_2023} that the boron dose in a certain point is proportional to boron concentration $C_{B}$ ($D_{B}\propto{}C_{B}$). The  measurement of real time in vivo boron concentration, necessary for dose determination, is one of the main challenges of BNCT. At present there is no real-time in vivo method to measure boron concentration during BNCT treatment, although several approaches are under investigation, the most promising being positron-emission tomography, magnetic resonance imaging and prompt gamma analysis. \emph{Prompt gamma analysis} with single photon emission tomography (SPECT) or Compton imaging systems aims to reconstruct boron distribution in vivo in real time using the prompt $478$ keV gamma rays emitted by the excited lithium nucleus in boron neutron capture reaction.

\subsection{Compton imaging}
\begin{figure}
    \centering
    \includegraphics[width=0.55\linewidth, keepaspectratio]{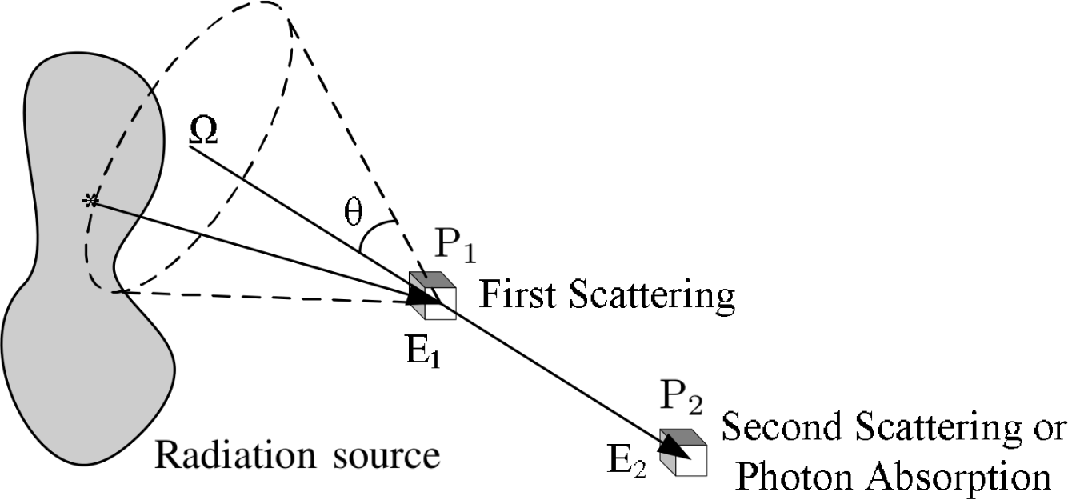}
    \caption[]{Schematic diagram of a general Compton camera.}
    \label{fig:compton_camera}
\end{figure}

A Compton camera is a gamma-ray detector that uses the kinematics of Compton scattering to reconstruct the original radiation source distribution.
\newline{}
The basic scheme of a Compton camera is shown in Fig.~\ref{fig:compton_camera}; The incoming gamma-ray undergoes Compton scattering from an electron in the position-sensitive first detector (scatterer) and then is absorbed by the second position-sensitive detector (absorber). The Compton scattering and absorption events are together called a \emph{Compton event}.
The measured data are the coordinates of the first and second point of interaction $\mathbf{d}_{1}=\left(X_{1},Y_{1},Z_{1}\right)$ and $\mathbf{d}  _{2}=\left(X_{2},Y_{2},Z_{2}\right)$, the energy $E_{1}$ deposited in the interaction of the scattered electron in the medium of the first detector and the energy $E_{2}$ deposited in the second detector. The energy of the incoming gamma-ray is equal to the the sum of the deposited energies: $E_{\gamma}=E_{1}+E_{2}$. Assuming that the electron before scattering has no momentum in the laboratory frame, from energy-momentum conservation it follows that the energy of the scattered photon as a function of the scattering angle in Compton scattering is:
\begin{equation}
\label{eq:2_energy_angle}
E_{\gamma}^{\prime}=\frac{E_{\gamma}}{1+\frac{E_{\gamma}}{m_{e}c^{2}}(1-\cos{\theta})},
\end{equation}
which allows the estimation of the Compton scattering angle $\theta$ from the energy deposited in the first detector and the gamma-ray energy. For each Compton event, the position of the source is confined to the surface of a cone, called \emph{Compton cone}, with the vertex in the interaction point $P_{1}$ in the scatterer, the axis passing also through the interaction point $P_{2}$ in the absorber, and the half-angle $\theta$ obtained from the energies measurements. It is possible to estimate the original gamma source distribution from the superimposition of Compton cones.
\par{}
While SPECT imaging is characterized by quite low sensitivity, due to the presence of the collimator, that determines a relation of inverse proportionality between the sensitivity and the spatial resolution squared  \citep{Nillius2008}, Compton cameras, which apply electronic collimation instead of physical collimation, are characterized by higher sensitivity; moreover sensitivity depends on the size, type and geometry of the two detectors, and is therefore independent of angular (and spatial) resolution, depending on the noise and spatial resolution characteristics of the detector. Another important characteristic of Compton imaging compared with SPECT is that angular (and spatial) resolution improves with increasing gamma ray energy, since from eq.~\ref{eq:2_energy_angle}:
\begin{equation}
    \label{eq:angle_uncertainty_energy}
    \dd{}\theta{}=\frac{m_{c^{2}}}{\sin{\theta}(E_{\gamma}-E_{1})^{2}}\dd{}E_{1}.
\end{equation}
On the other hand, in SPECT the sensitivity decreases for higher energies, since the septal thickness must be increased to reduce gamma ray penetration. Furthermore, Compton imaging could in principle allow the rejection of most of the background events due to \textsuperscript{10}B present in the shielding walls of most BNCT facilities, by filtering out all Compton cones having zero intersection with the reconstruction region.

\subsection{Compton image reconstruction}
A general imaging system is associated with a linear integral operator $\mathcal{H}$ such that the continuous source distribution $f(\mathbf{x})$ and the expected value of the projection measurements $g(\mathbf{y})$ are related by:
\begin{equation}
\label{eq:imaging_system_integral}
g(\mathbf{y}) = (\mathcal{H}f)(\mathbf{y}) = \int_{\mathcal{X}}h(\mathbf{y}, \mathbf{x})f(\mathbf{x})\dd{\mathbf{x}}
\end{equation}
or more precisely $g(\mathbf{y}) = (\mathcal{H}f)(\mathbf{y})+\epsilon$, if noise is taken into account\footnote{The kernel $h(\mathbf{y}, \mathbf{x})$, representing the conditional probability of detecting an event at location $\mathbf{y}$ given that it was emitted at location $\mathbf{x}$ in the source, is also called \emph{space-variant (projection) point spread function} \citep{bertero_inverse, Chen_Wei_Xue_2004}, since it amounts to the response of the system to a point source, that is to a Dirac delta distribution. }. In the case the case of Compton imaging the source coordinate $\mathbf{x}=[x,y,z]$ in equation~\ref{eq:imaging_system_integral} represents a 3D location, while the detector coordinate $\mathbf{y}=[\mathbf{d}_{1},\mathbf{d}_{2},E_{1},E_{2}]$ is a vector containing the Compton event detection locations and the energies deposited in the two detectors.
\par{}
Since the projection operator $\mathcal{H}$ is a compact operator and in particular a \emph{Hilbert-Schmidt operator}\footnote{The supports $\mathcal{X}$ and $\mathcal{Y}$ are finite and the kernel $h(\mathbf{y}, \mathbf{x})$ is bounded for all $\mathbf{x}$ and $\mathbf{y}$, thus $\int_{\mathcal{X}}\int_{\mathcal{Y}}h^{2}(\mathbf{y}, \mathbf{x})\dd{\mathbf{x}}<+\infty$. A Hilbert-Schmidt operator can be approximated with arbitrary accuracy, in the Hilbert-Schmidt norm, by a finite rank operator \citep{van_neerven}.} \citep{van_neerven,WA04_emiss_tomogr}, if we consider the discrete approximation $\mathbf{f}=[f_{1}, ..., f_{B}]^{T}$ of the source distribution (pixel values) and $\mathbf{g}=[g_{1}, ..., g_{D}]^{T}$ is the expected number of events in each of the $D$ detector channels, the \emph{discrete} version of eq.~\ref{eq:imaging_system_integral} is obtained:
\begin{equation}
\label{eq:imaging_system_discrete}
\mathbf{g} = \mathbf{H}\mathbf{f} 
\end{equation}
where $\mathbf{H}$ is a $D\times{}B$ matrix called \emph{system matrix}.
\par{}
The aim of image reconstruction is to determine an estimate $\hat{f}(\mathbf{x})$ of $f(\mathbf{x})$ given the noisy measurements and the operator $\mathcal{H}$.\par{}
\par{}
Traditionally, the \emph{analytic approach} has been employed to solve the inverse problem by neglecting any explicit randomness and deterministic blurring and attenuation mechanisms and trying to exactly invert eq.~\ref{eq:imaging_system_integral} with a simple enough projection operator $\mathcal{H}$. On the other hand modern reconstruction techniques employ more general linear models that can take into account the blurring and attenuation mechanisms and generally also incorporate probabilistic models of noise\footnote{A reconstructed image is affected by both \emph{determinic} degradations, like \emph{blur}, linked to the point spread function of the imaging procedure, and \emph{aliasing}, caused by sampling, and \emph{random} degradations, collectively referred to as noise \citep{jain,gonzalez,prob_random_statistical}.} without requiring restrictive system geometries \citep{WA04_emiss_tomogr}. These models typically do not admit an explicit solution, or an analytic solution can be difficult to compute, so in most cases these models are dealt with by \emph{iterative algorithms}, in which the reconstructed image is progressively refined in repeated calculations. In this way a greater accuracy can be obtained, although longer computation times are required.
\par{}
The most commonly employed algorithms for Compton image reconstruction are MLEM and MAP. The maximum likelihood (ML) criterion is the most used technique in statistical inference for deriving estimators. In this criterion it is presumed that the observation vector $\mathbf{g}$ is determined by an unknown deterministic parameter $\mathbf{f}$, which in this case is the gamma source distribution to be reconstructed, following the conditional probability $p(\mathbf{g}|\mathbf{f})$. The \emph{maximum likelihood estimate} $\hat{\mathbf{f}}$ of $\mathbf{f}$ is the reconstructed image maximizing the likelihood function $\mathfrak{L}(\mathbf{f})=p(\mathbf{g}|\mathbf{f})$ for the measured data $\mathbf{g}$:
\begin{equation}
    \label{ML_estimate}
    \hat{\mathbf{f}}=\underset{\mathbf{f}}{arg\,max}\;p(\mathbf{g}|\mathbf{f}).
\end{equation}
The \emph{maximum likelihood expectation maximization} (MLEM) algorithm is an iterative procedure whose output tends to the ML solution of the problem. In the case of inversion problems with Poisson noise, it consists in the following iteration step \citep{WA04_emiss_tomogr}:
\begin{equation}
    \label{MLEM_step}
    \hat{f}_{j}^{(n+1)}=\frac{\hat{f}_{j}^{(n)}}{\sum\limits_{i}^{}h_{ij}}\sum\limits_{i}^{}h_{ij}\frac{g_{i}}{\sum\limits_{k}^{}h_{ik}\hat{f}_{k}^{(n)}}.
\end{equation}
The two main shortcomings of MLEM reconstruction are the slow convergence (typically $30-50$ iterations are required to get usable images) and high noise.
\newline
The iterative step in eq.~\ref{MLEM_step}, also known as \emph{binned-mode} MLEM could be applied to reconstruct the image. However in the case of a Compton camera the number $D$ of possible detector elements in $\mathbf{g}$ can be exceedingly high, for example a practical Compton camera could have $2^{16}$ first-detector elements, the same number of second-detector elements and $2^{8}$ energy channels, resulting in $D\approx{}10^{12}$ \citep{WA04_emiss_tomogr}. Since a typical Compton camera dataset size is of the order of $10^{8}$ events, most of the possible detector bins will contain zero. An alternative approach, the so called \emph{list-mode} MLEM \citep{ VALENCIALOZANO2023124, Maxim_2016, Wilderman_773871, Parra_873014} can be adopted in order to reduce computational cost. In this approach the number of detector bins $D$ is assumed to be very large, so that most of them contain zero counts, while the occupied bins contain only one count. This is basically equivalent to consider infinitesimal bins. If $N$ is the number of detected events, $\mathcal{D}=\{d_{1},\dots{},d_{N}\}$ denotes the set of all occupied bins and the  observation vector $\mathbf{g}$ becomes:
\begin{equation}
    g_{d} =
    \begin{cases*}
      1 & if  $d\in\mathcal{D}$\\
      0 & otherwise
    \end{cases*}.
\end{equation}
The list mode iteration is then given by:
\begin{equation}
    \label{list_mode_MLEM_step}
    \hat{f}_{b}^{(n+1)}=\frac{\hat{f}_{b}^{(n)}}{s_b}\sum\limits_{d\in\mathcal{D}}^{}h_{db}\frac{g_{d}}{\sum\limits_{b}^{}h_{db}\hat{f}_{b}^{(n)}},
\end{equation}
where $s_{b}=\sum\limits_{d}^{}h_{db}>\sum\limits_{d\in{}\mathcal{D}}^{}h_{db}$ is the sum over all detector bins (and not only on the occupied ones) of the probabilities $h_{db}$ of the gamma ray emitted from the voxel $b$ being detected in the detector bin $d$. $s_{b}$ therefore represents the probability that of a gamma ray emitted from $b$ would be detected and is called \emph{sensitivity}. As a first approximation $s_{b}$ can be considered equal to the solid angle subtended by the scatter detector at bin $b$ divided by $4\pi{}$, which becomes essentially uniform over all the voxels if the detector is small \citep{Maxim_2016}. The calculation of the expressions of the system matrix and sensitivity is carried out in \citep{Maxim_2016}. Using equation~\ref{list_mode_MLEM_step} the sum ranges only over the number of events rather than all detector elements, and computational cost is thus generally reduced.
\par{}
\subsection{Research outline}
In order to investigate the potentialities of Compton imaging with CZT detectors for BNCT, a Geant4 simulation of a simplified detector in a BNCT setting has been implemented. The data from the simulation has been used to reconstruct gamma sources distribution with the list-mode MLEM algorithm.
\par{}
Models based on deep neural networks for reconstructing the dose distribution from the simulated dataset of BNCT Compton camera images while avoiding the long iteration time associated with the MLEM algorithm have been examined.
\par{}
The U-Net architecture and two variants based on the deep convolutional framelets approach have been used for noise and artifacts reduction in few-iterations reconstructed images, leading to promising results in terms of reconstruction accuracy and processing time.

\section{Deep learning models}

The past few years have witnessed impressive advancements in the application of deep learning to biomedical image reconstruction. While classical algorithms generally perform well, they often require long processing times. The use of deep learning techniques can lead to orders of magnitude faster reconstructions in some cases even with better image quality than classical iterative methods \citep{Ben_Yedder_2020_dl_reco_survey, dl_inverse_problems, YE_deep_biomedical_reco}.
\par{}
There exist various approaches to apply deep learning to solve the inverse problem. The taxonomy of deep learning approaches to solve inverse problems, based on a first distinction between supervised and unsupervised techniques and a second distinction considering what is known and when about the forward model $\mathbf{H}$, consists of sixteen major categories described in \citep{dl_inverse_problems}.
\par{}
The present study focuses on the case of supervised image degradation reduction with U-Nets in the deep convolutional framelets framework, with forward operator $\mathbf{H}$ fully known, using a matched dataset of measurements $\mathbf{g}$, the Compton events data produced in a Geant4 simulation and ground truth images $\mathbf{f}$, corresponding to the output of the $60$th iteration of the list-mode MLEM algorithm \citep{VALENCIALOZANO2023124,Maxim_2016} applied to the Compton events data (reconstruction time $\approx{}24-36$ minutes). The objective in the supervised setting is to obtain a \emph{reconstruction network} $r_{\theta}(\cdot{})$ mapping measurements $\mathbf{g}$ to images $\hat{\mathbf{f}}$, where $\theta$ is a vector of parameters to be learned.

\subsection{Image degradation reduction: U-Nets and deep convolutional framelets}
\label{subsec:unets}
\begin{figure}[htbp]
    \centering
    \includegraphics[width=0.6\linewidth]{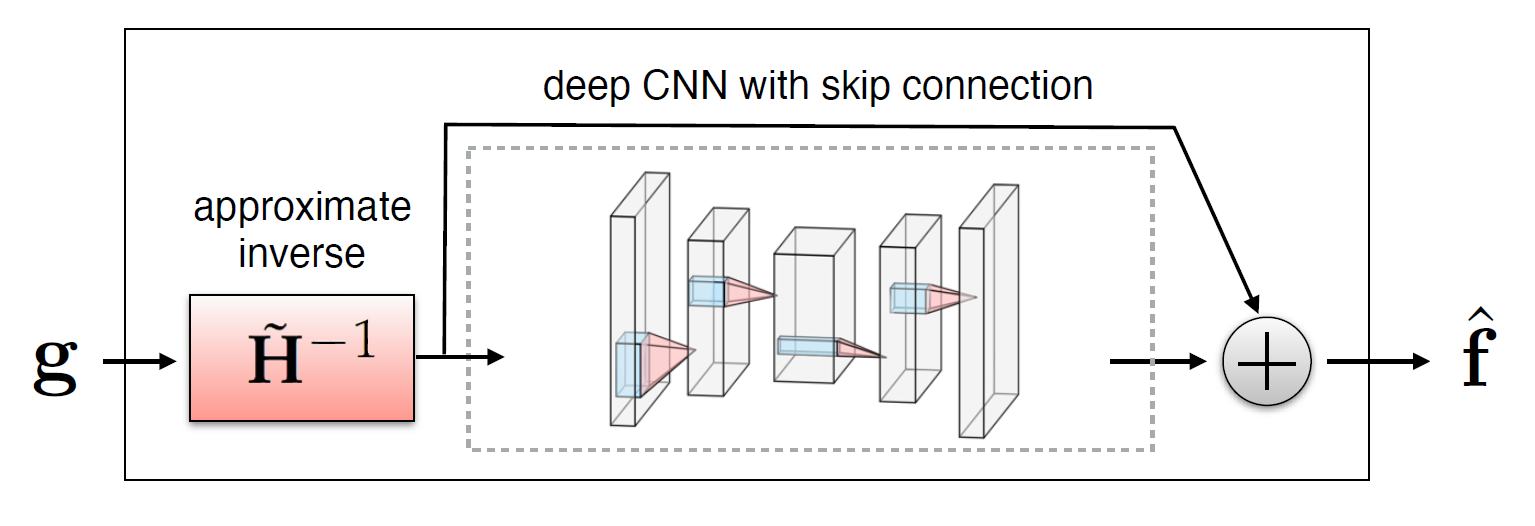}
    \caption[]{Illustration of a CNN with skip connection to remove noise and artifacts from an initial reconstruction obtained by applying $\tilde{\mathbf{H}}^{-1}$ to measurements \citep{dl_inverse_problems}.}
    \label{degradation_reduction_approach}
\end{figure}

A simple method for embedding the forward operator $\mathbf{H}$ into the network architecture is to apply an approximate inverse operator $\tilde{\mathbf{H}}^{-1}$\footnote{A matrix such that $\tilde{\mathbf{H}}^{-1}\mathbf{H}\mathbf{f}\approx{}\mathbf{f}$ for every image $\mathbf{f}$ of interest} to first map measurements to image domain and then train a neural network to remove degradations (noise and artifacts) from the resulting images \citep{dl_inverse_problems}, as illustrated in Fig.~\ref{degradation_reduction_approach}.
The expression of the reconstruction network is therefore:
\begin{equation}
    \hat{\mathbf{f}}=r_{\theta{}}(\mathbf{g})=n_{\theta{}}(\tilde{\mathbf{H}}^{-1}\mathbf{g})+\tilde{\mathbf{H}}^{-1}\mathbf{g},
\end{equation}
where $n_{\theta{}}$ is a trainable neural network depending on parameters $\theta{}$. Networks with more complicated hierarchical skip connections are also commonly used. In this study  the use of the standard U-Net architecture and two variants satisfying the so-called frame condition, the dual frame U-Net and the tight frame U-Net, first proposed in \citep{ye_framing_unet}, is examined.
\par{}
The \emph{deep convolutional framelets} framework, introduced by Ye in \citep{deep_convolutional_framelets}, provides a theoretical understanding of convolutional encoder-decoder architectures, such as U-Nets, by adopting the frame-theoretic viewpoint \citep{casazza_13}, according to which the forward pass of a CNN can be regarded as a decomposition in terms of a frame that is related to pooling operations and convolution operations with learned filters \citep{math_dl_22,ye_geometry_22}. Deep convolutional framelets are characterized by an inherent \emph{shrinking behavior} \citep{deep_convolutional_framelets}, which determines the degradation reduction capabilities allowing the solution of inverse problems.
\par{}
\begin{figure}[htbp]
	\centering
        \subfloat[][]{\includegraphics[width=0.75\linewidth]{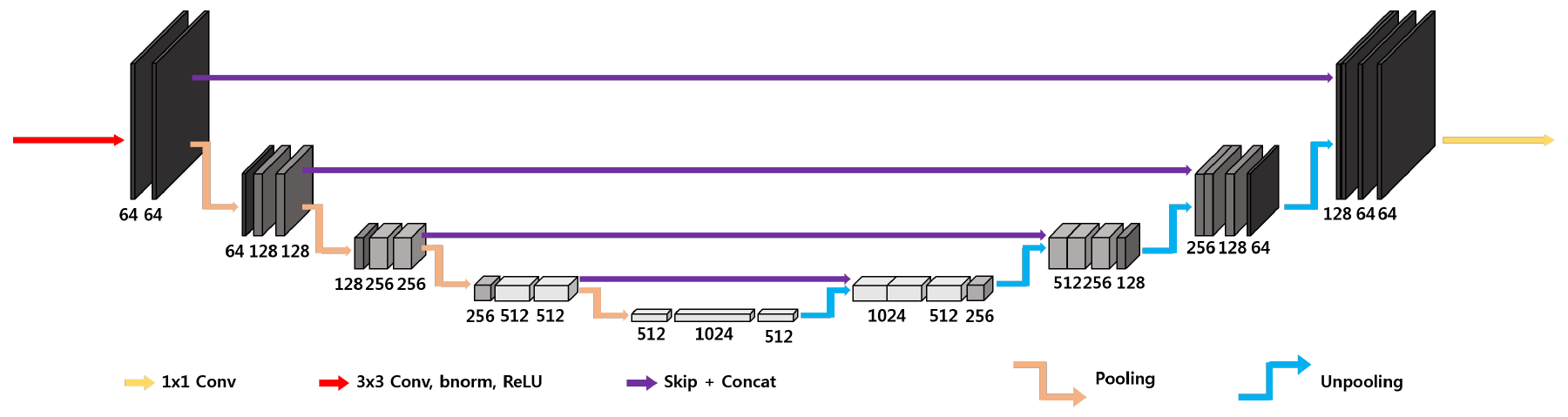}\label{fig_unet}} \quad
	\subfloat[][]{\includegraphics[width=0.75\linewidth]{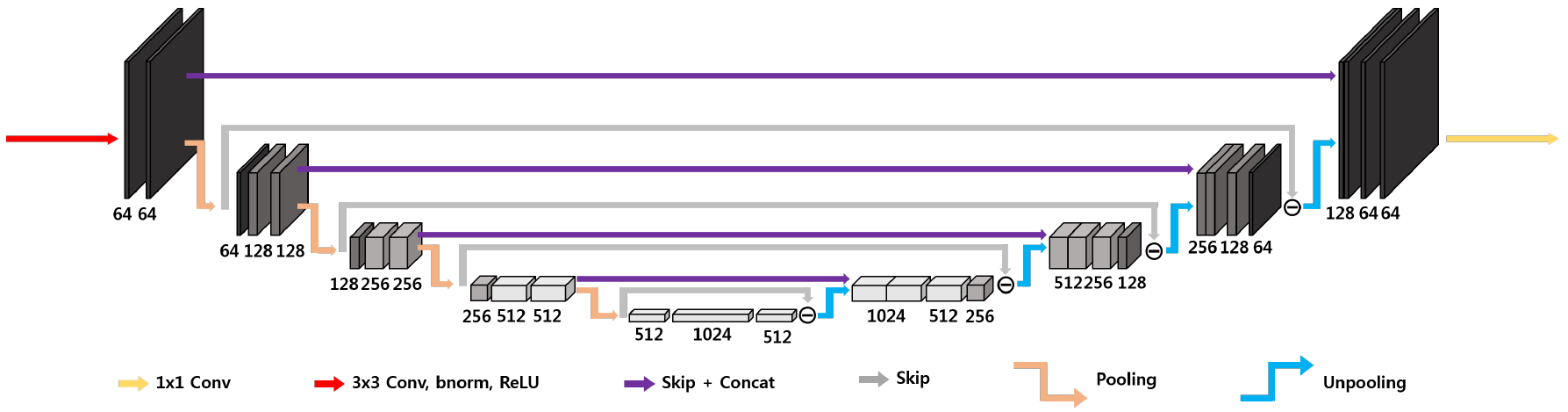}\label{fig_dual}} 
	\caption{Simplified 3D architecture of (a) standard U-Net, (b) dual frame U-Net \citep{ye_framing_unet}. These are 4D representations, the plane perpendicular to the page corresponds three-dimensional space.}
	\label{fig:architectures_3D}
\end{figure}
The three-dimensional U-Net architecture, illustrated in Fig.~\ref{fig_unet}, initially proposed for biomedical image segmentation, is widely used for inverse problems \citep{unser_deep_convolutional_inverse}. The network is characterized by an encoder-decoder structure organized recursively into several levels, with the next level applied to the low-resolution signal of the previous layer \citep{ye_framing_unet}; the encoder part consists of $3\times{}3$ convolutional layers, average pooling layers, denoted by $\Phi{}^\top$, batch normalization and ReLUs, and the decoder consists of average unpooling layers, denoted by $\Phi{}$, and $3\times{}3$ convolution. There are also skip connections through channel concatenation, which allow to retain the high-frequency content of the input signal. The pooling and unpooling layers determine an exponentially large receptive field. As outlined in \citep{ye_framing_unet}, the extended average pooling and unpooling layers ($\Phi_{ext}^\top := \begin{bmatrix} I   \\ \Phi^\top \end{bmatrix}$ and $\Phi_{ext}=\begin{bmatrix} I & \Phi{} \end{bmatrix}$, respectively) do not satisfy the frame condition, which leads to an overemphasis of the low frequency components of images due to the duplication of the low frequency branch \citep{deep_convolutional_framelets}, resulting in artifacts.
\par{}
A possible improvement is represented by the three-dimensional dual frame U-Net, employing the dual frame of $\Phi_{ext}$, given by \citep{ye_framing_unet}:
\begin{equation}
    \tilde \Phi_{ext} =  (\Phi_{ext}\Phi_{ext}^\top)^{-1}\Phi_{ext}=  \begin{bmatrix} I-\Phi\Phi^\top/2 & \Phi/2 \end{bmatrix},
\end{equation}
corresponding to the architecture in Fig.~\ref{fig_dual}. In this way the frame condition is satisfied, but there exists noise amplification linked to the condition number of $I+\Phi\Phi^\top$, which is equal to $2$ \citep{ye_framing_unet}.
\par{}
\begin{figure}
    \centering
    \includegraphics[width=0.6\linewidth, keepaspectratio]{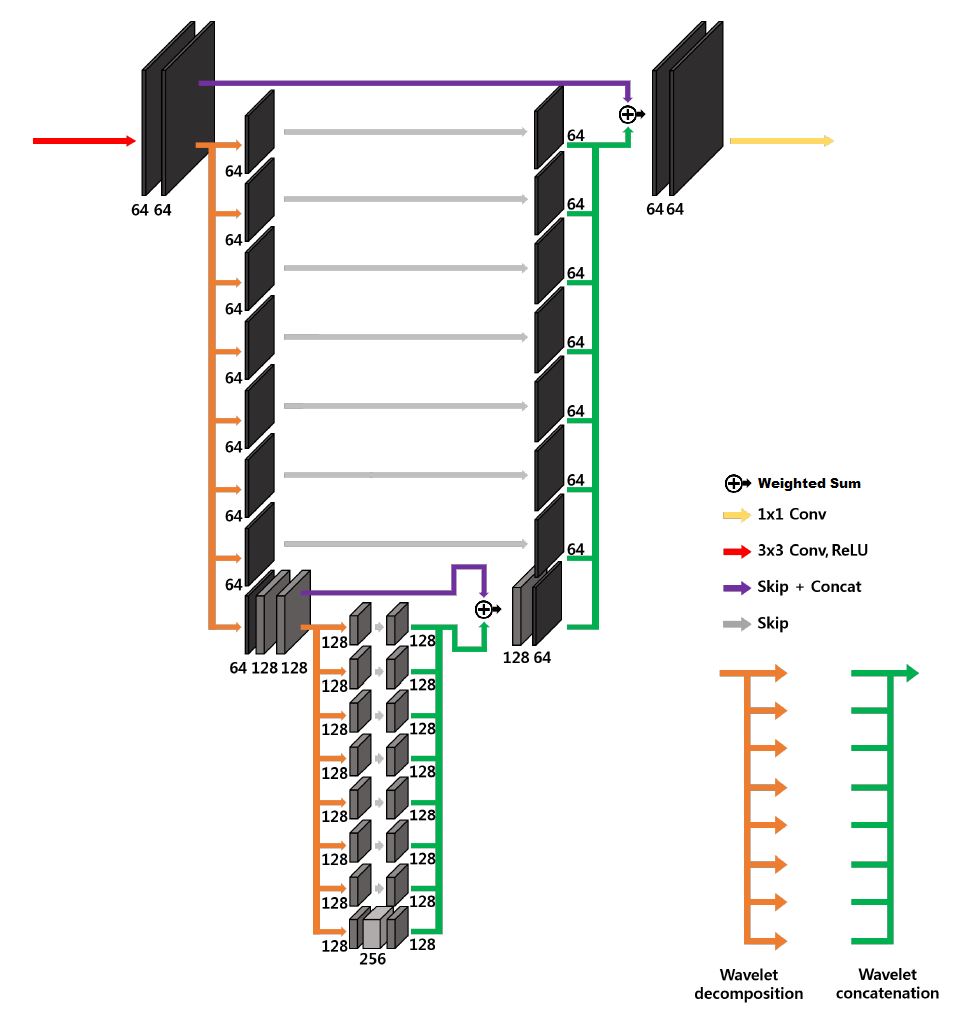}
    \caption[]{Modified 3D tight frame U-Net. This is a 4D representation, the plane perpendicular to the page corresponds three-dimensional space.}
    \label{fig:tight_frame_3D}
\end{figure}
The usage of tight filter-bank frames or wavelets allows to improve the performance of the U-Net by satisfying the frame condition with minimum noise amplification. In this case the non-local basis $\Phi^\top$ is now composed of a tight filter-bank:
\begin{equation}
\Phi^\top = \begin{bmatrix} T_1^\top  & \cdots & T_L^\top \end{bmatrix}^\top,
\end{equation}
where $T_k$ denotes the $k$-th subband  operator. The simplest tight filter-bank frame is the Haar wavelet transform \citep{mallat,damelin_mathematics_signal}. In $n$ dimensions it is composed by $2^{n}$ filters.  
The three-dimensional tight frame U-Net architecture is in principle essentially analogous to the one-dimensional and two-dimensional ones proposed in \citep{ye_framing_unet}. The main difficulty is linked to the fact that operations on three-dimensional images are computationally more expensive and the number of filters in the filter bank rises to $8$. In order to reduce the computational cost, the large-output signal concatenation and multi-channel convolution have been substituted by a simple weighted sum of the signals with learnable weights, as shown in Fig.~\ref{fig:tight_frame_3D}, representing a two-level 3D modified tight frame architecture. Notice that this substitution drastically reduces the number of parameters to be learnt, reducing the possibility of overfitting.

\section{Methods}

\subsection{Monte Carlo simulation}
Since Compton imaging requires detectors characterized by both high spatial and energy resolutions \citep{Tashima2022-mb,WA04_emiss_tomogr}, high resolution 3D CZT drift strip detectors, which currently offer the best performance \citep{Abbene_recent_advances,Abbene_potentialities_BNCT}, were considered in this study.
\par
\begin{figure}[htbp]
	\centering
        \subfloat[][]{\includegraphics[width=0.33\linewidth]{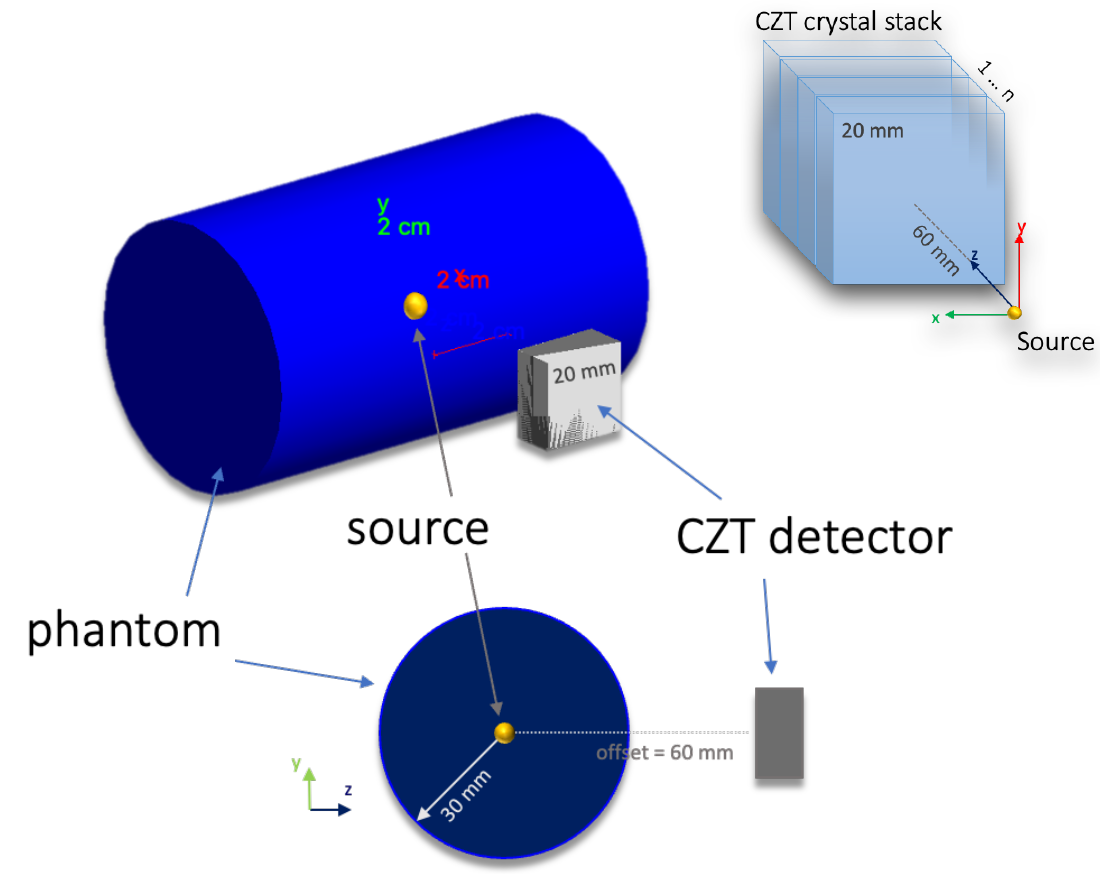}\label{original_det}} \quad
	\subfloat[][]{\includegraphics[width=0.3\linewidth]{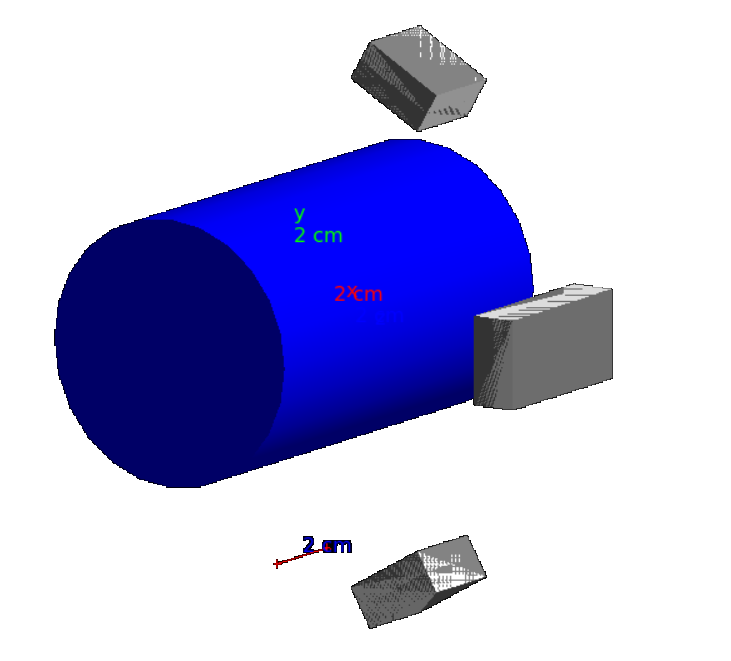}\label{44det}} \quad
	\subfloat[][]{\includegraphics[width=.26\linewidth]{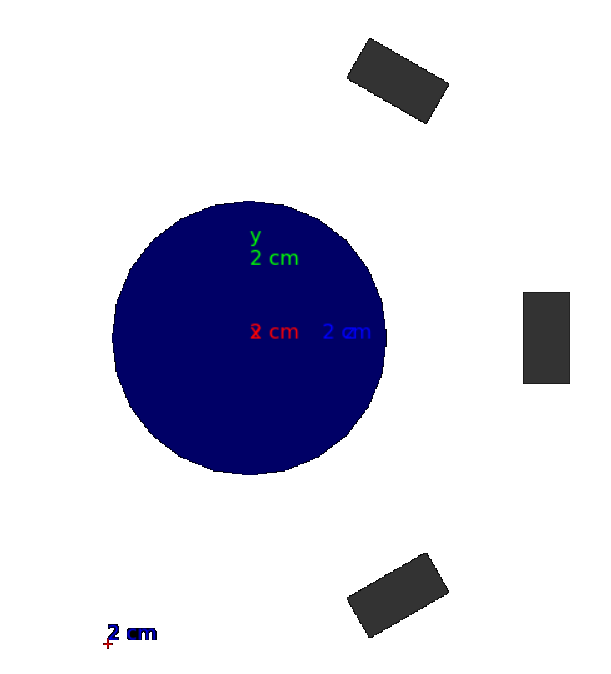}\label{44det_YZ}}
	\caption{(a) Single module geometry, (b) four modules geometry and (c) four modules geometry YZ view.}
	\label{fig:det}
\end{figure}

{}
A Geant4 simulation of the CZT detector modules in a BNCT setting was implemented in order to assess the accuracy in the source reconstruction and optimize detector geometry. For the simulation the G4EMStandardPhysics and G4DecayPhysics classes were used. The range cut value of $0.7$ $mm$ was set. A small animal (or human patient irradiated body part with comparable dimensions) was simulated with a cylinder of radius $30$ $mm$ and height $100$ $mm$, declared with the standard soft tissue material G4\_TISSUE\_SOFT\_ICPR. A single 3D CZT sensor module is composed by four 3D CZT drift strip detectors described in \citep{Abbene_recent_advances}, each of which  ideally assumed to be a CZT $20$ $mm\times{}20$ $mm\times{}5$ $mm$ parallelepiped in the simulation, stacked along the $y$ direction (see Fig.~\ref{original_det}). Besides the configuration with a single sensor module with the $X_{det}Z_{det}$ plane parallel to the cylinder axis, placed at a distance of $60mm$ from the cylinder center, equidistant from the cylinder bases, illustrated in Fig.~\ref{original_det}, other configurations with different numbers of modules in different positions were considered. 
\par{}
A good configuration in terms of source reconstruction quality and cost was found to be the four modules configuration in Fig.~\ref{44det}-\ref{44det_YZ}. 
In the four modules configuration two modules are obtained by applying a rotation of $\pm{}60^{\circ}$ around the cylinder axis to the original single module of the previous configuration, and with the other two modules obtained by translating by $\pm{10}$ $mm$ the original module along the cylinder axis, so as to obtain a larger effective module given by the union of the two. An idealized case ha been considered with respect to the real BNCT energy spectrum \citep{Abbene_potentialities_BNCT}: all gamma rays are generated with the same $478$ $keV$ energy, the energy of the gamma rays emitted in the boron neutron capture reaction. The gamma emission is set to be isotropic.
\par{}
Since different tumor region geometries are needed for the training phase of deep learning algorithms, $20$ different tumor region geometries were created. For $17$ of these $3:1$, $4:1$, $5:1$ and $\infty{}$ (zero boron concentration in normal tissue) concentration ratios have been considered, while for the other three only the $\infty{}$ concentration ratio has been considered, for a total of $71$ different gamma sources distributions. Moreover, for each of these, $4$ different roto-translations of the tumor region were created, obtaining $355$ different different gamma sources distributions. 
\par{}
In order to select the events of interest, a filter was applied to consider only events with a total energy deposition in the range $0.470$ $MeV-0.485$ $MeV$ and with interaction points in the detector crystals. The simulation returns for each event the gamma source position, Compton scattering interaction position, photoabsorption interaction position, Compton recoil electron deposited energy and the photoabsorption deposited energy.
\par{}
For each possible gamma sources distribution $300$ million events were generated. About $1$ million events were detected by the apparatus in each run.
\par{}
\begin{figure}[htbp]
	\centering
	\subfloat[][]{\includegraphics[width=0.4\linewidth]{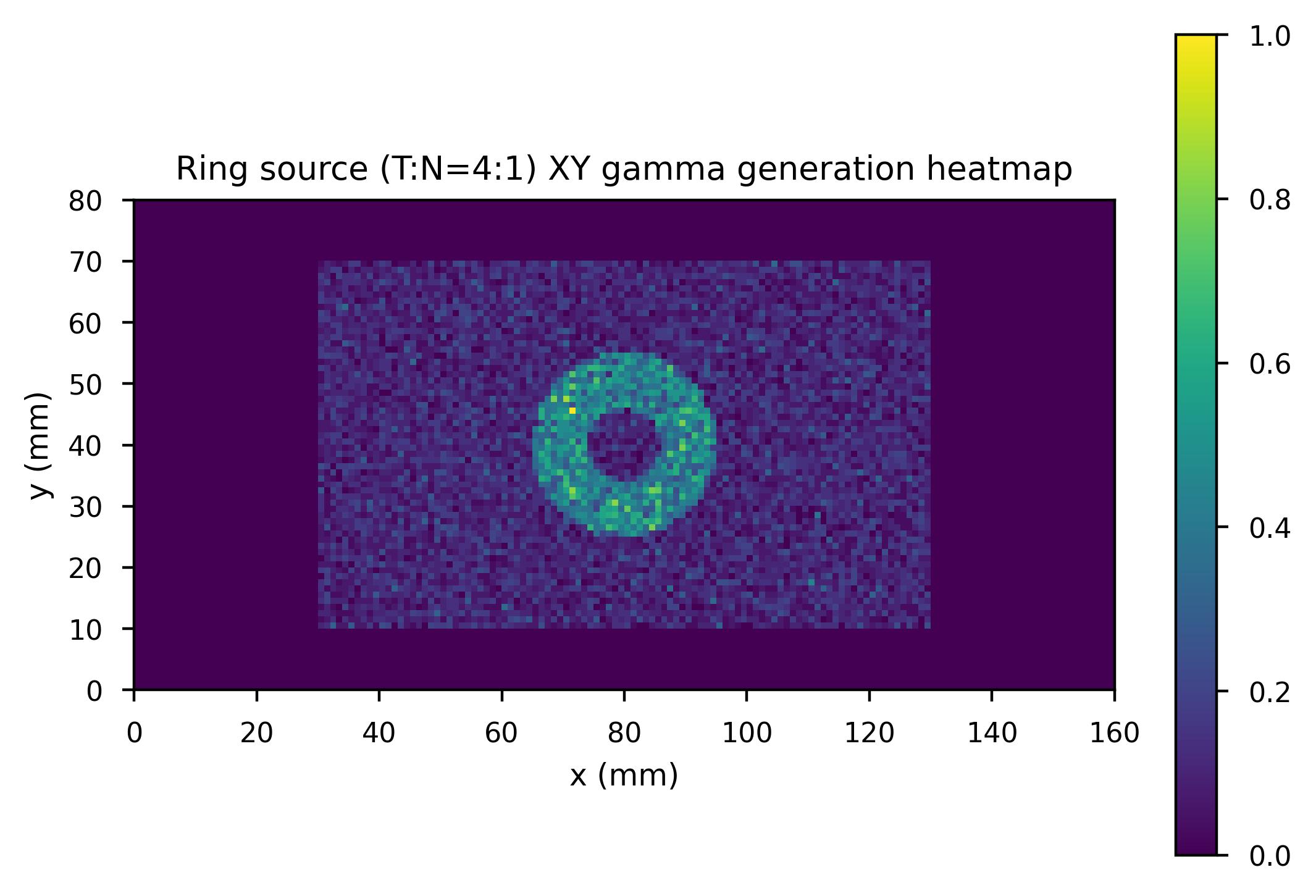}\label{prim8_frontal}} \quad
	\subfloat[][]{\includegraphics[width=0.4\linewidth]{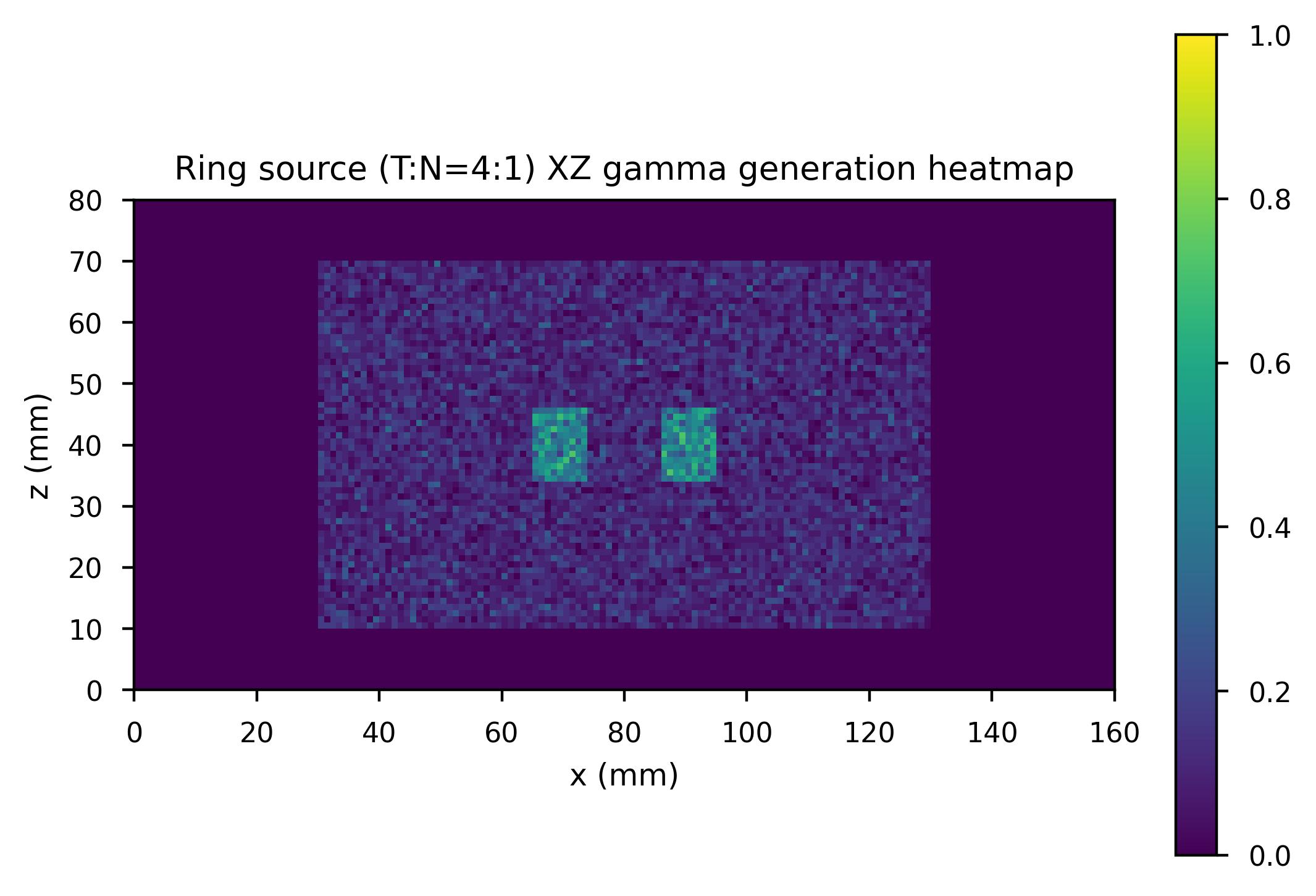}\label{prim8_XZ}} \quad
    \subfloat[][]{\includegraphics[width=0.40\linewidth]{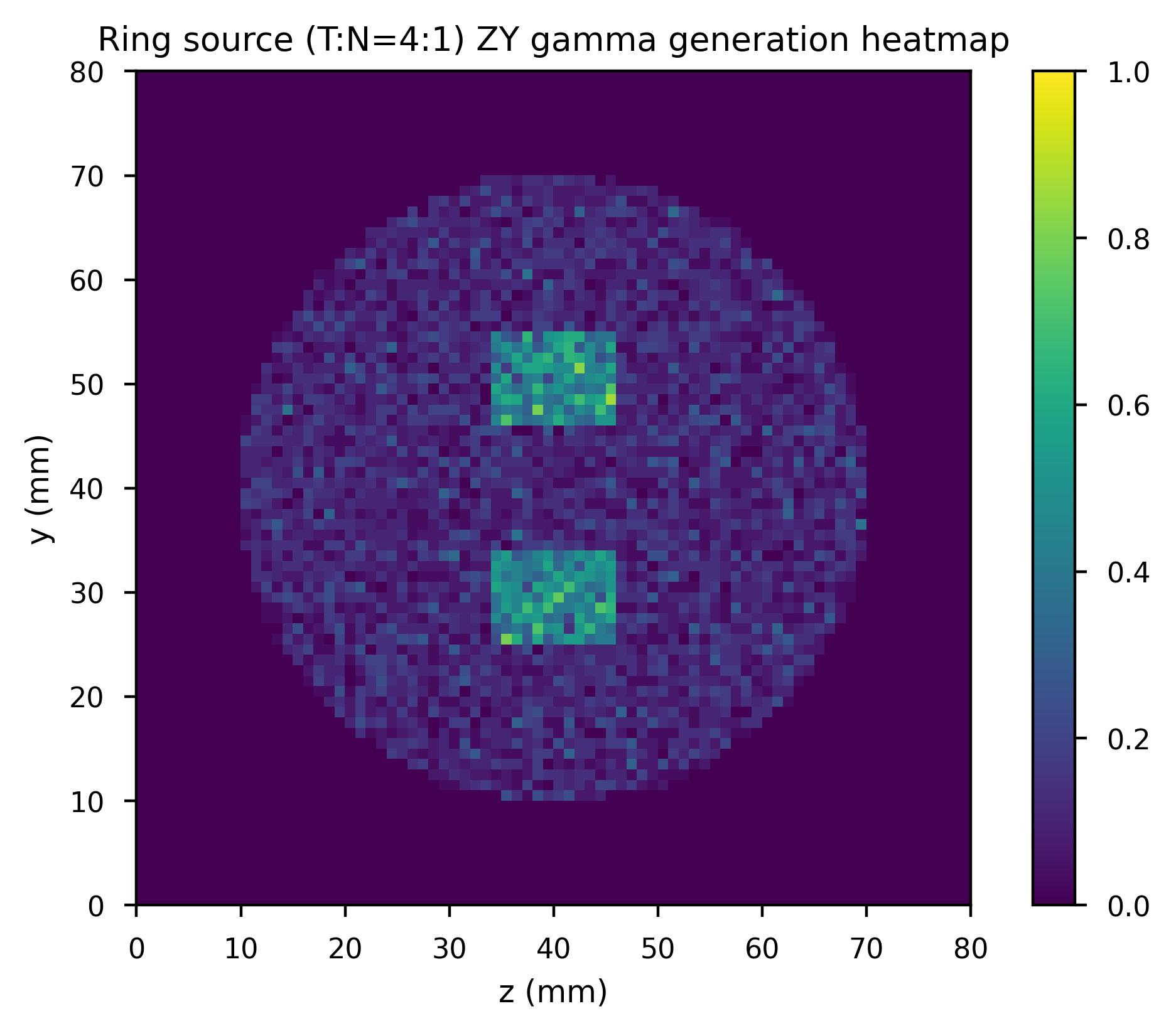}\label{prim8_YZ}} \quad
	\subfloat[][]{\includegraphics[width=.4\linewidth]{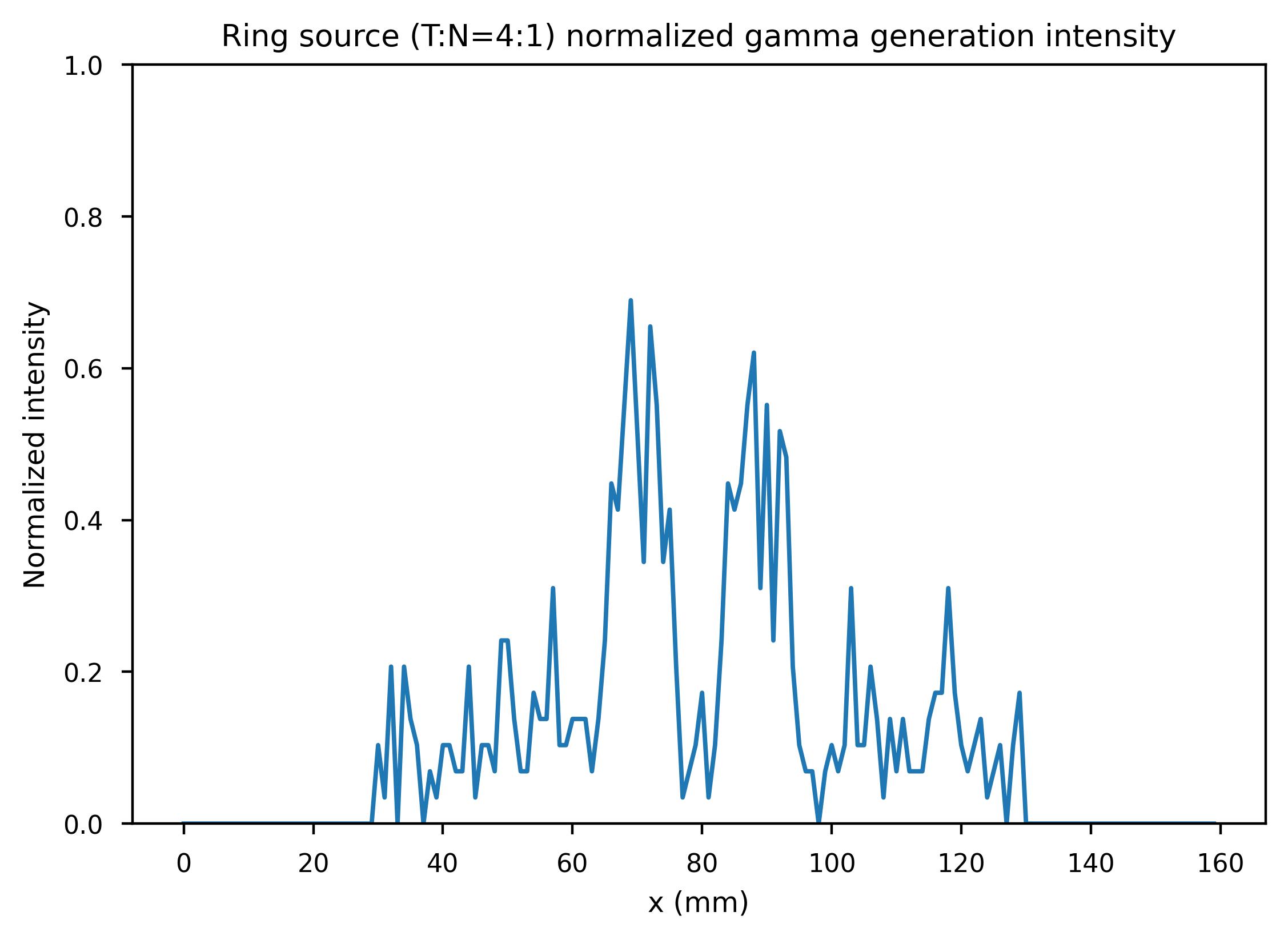}\label{prim8_1D}}
	\caption{Ring source, $T/N=4:1$ (a) XY gamma generation heatmap, (b) XZ gamma generation heatmap, (c) YZ gamma generation heatmap and (d) normalized intensity as a function of $x$.}
	\label{fig:prim8}
\end{figure}
Fig.~\ref{fig:prim8} shows the ring gamma source distribution with a concentration ratio $4:1$ in the parallelepipedonal region of size $160$ $mm\times{}80$ $mm\times{}80$ $mm$. Fig.~\ref{prim8_frontal} is the XY gamma generation heatmap for $z=40$ $mm$, Fig.~\ref{prim8_XZ} represents the XZ gamma generation heatmap for $y=40$ $mm$, Fig.~\ref{prim8_YZ} is the YZ gamma generation heatmap for $x=80$ $mm$ and Fig.~\ref{prim8_1D} the normalized intensity as a function of $x$ for $y=z=40$ $mm$. 

\subsection{U-Nets: dataset, network architectures, training and evaluation}
\begin{figure}[htbp]
	\centering
	\subfloat[][]{\includegraphics[width=0.4\linewidth]{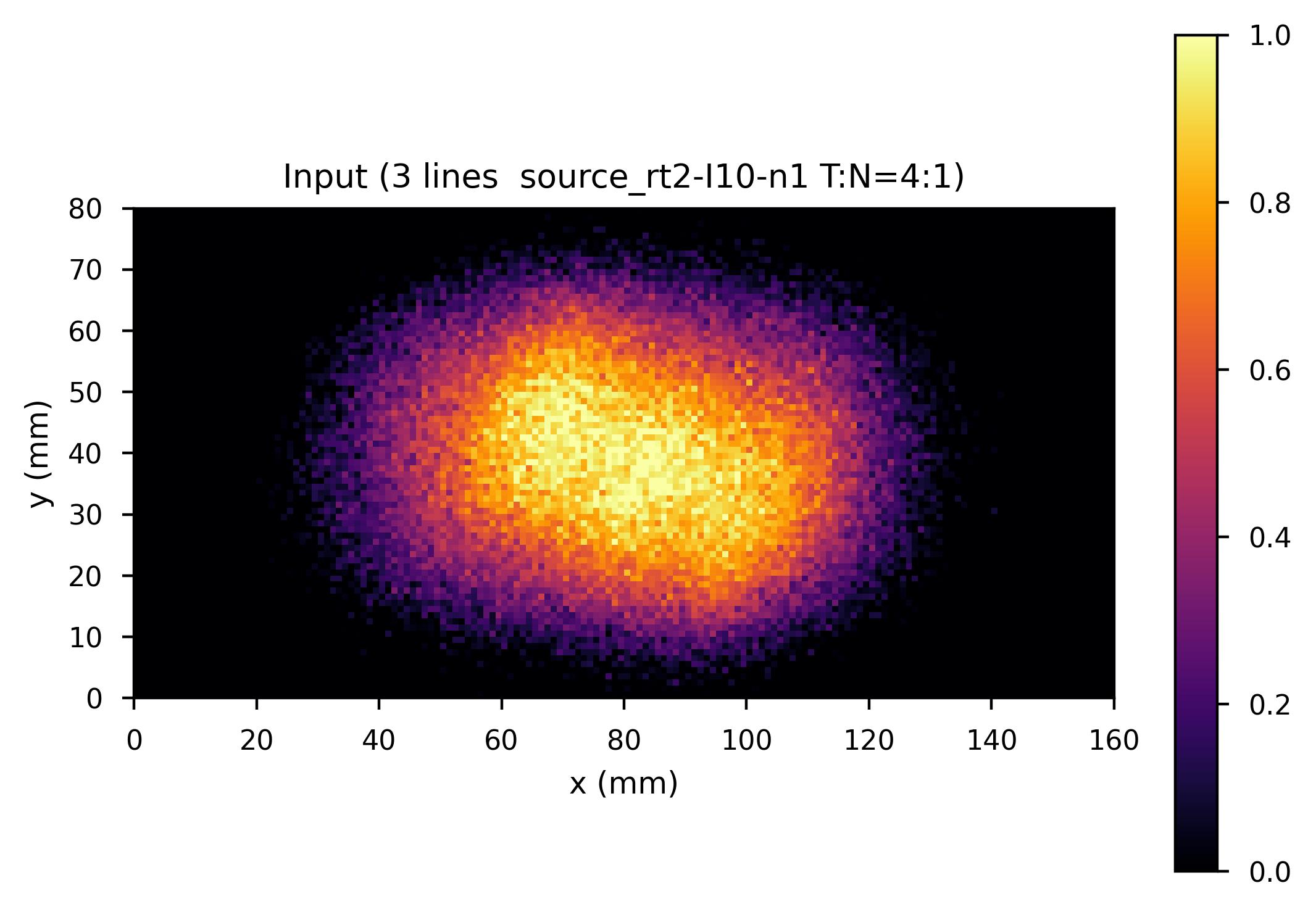}\label{input_0}} \quad
    \subfloat[][]{\includegraphics[width=0.4\linewidth]{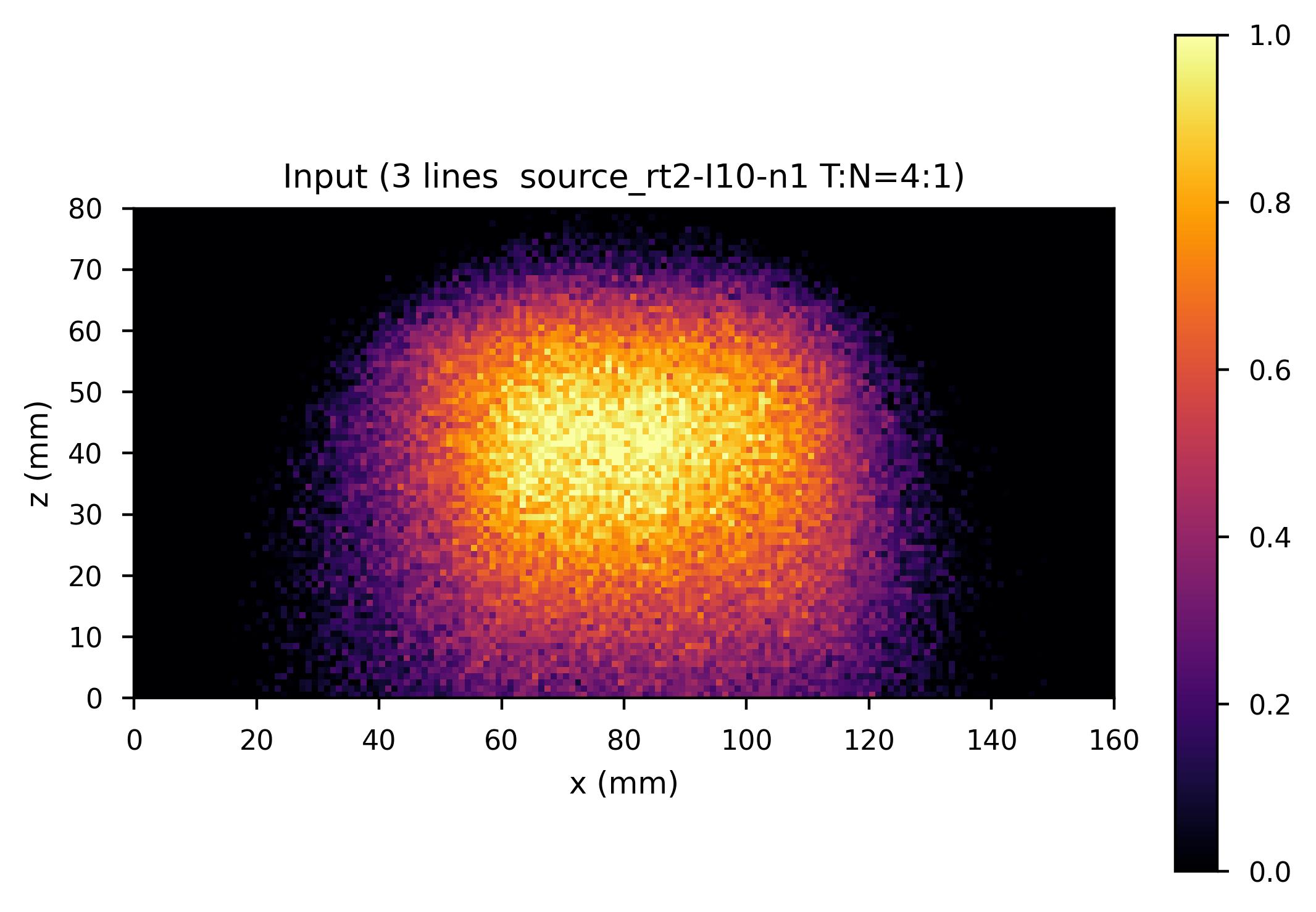}\label{input_1}} \quad
    \subfloat[][]{\includegraphics[width=0.4\linewidth]{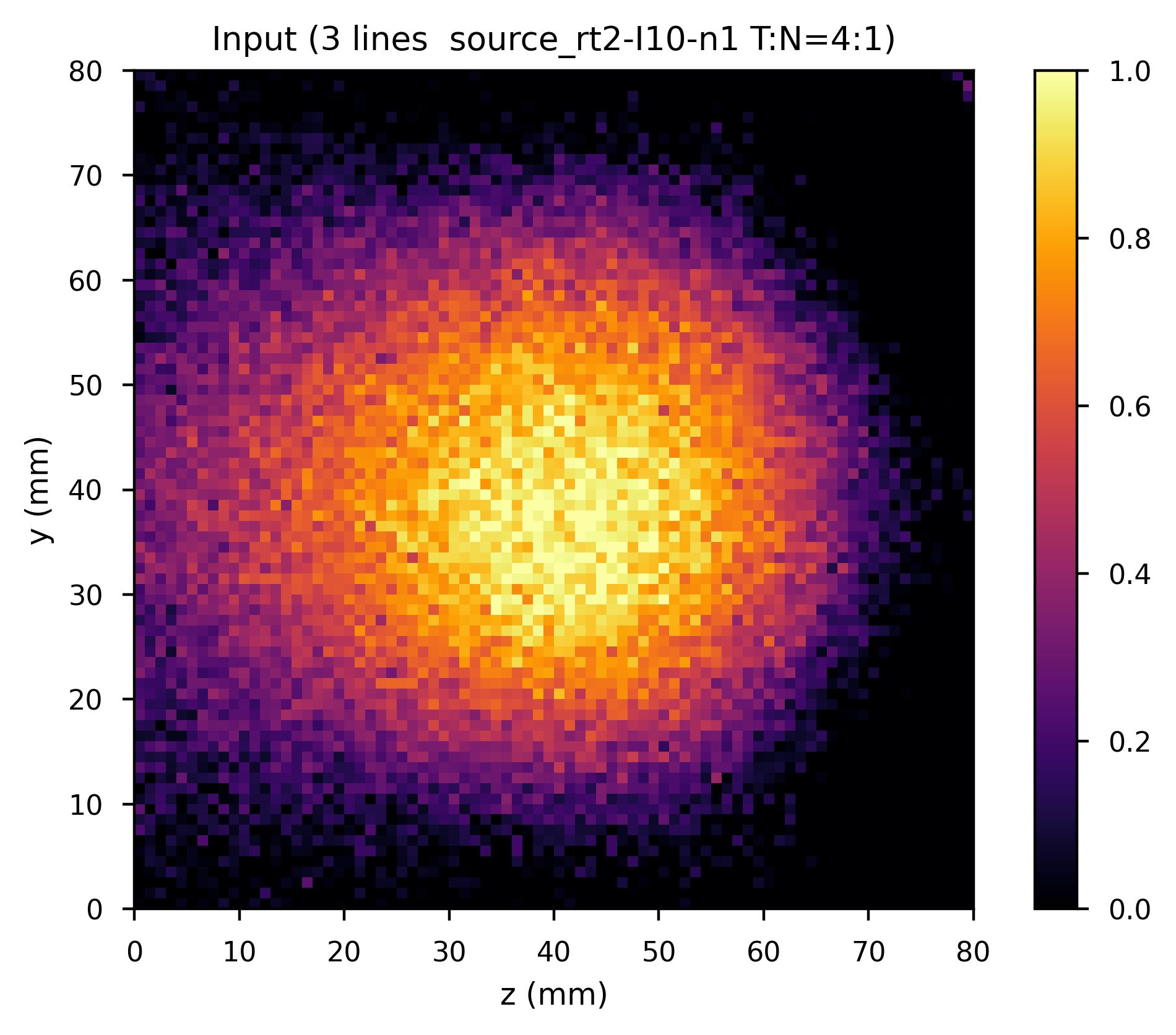}\label{input_2}} \quad
	\subfloat[][]{\includegraphics[width=0.4\linewidth]{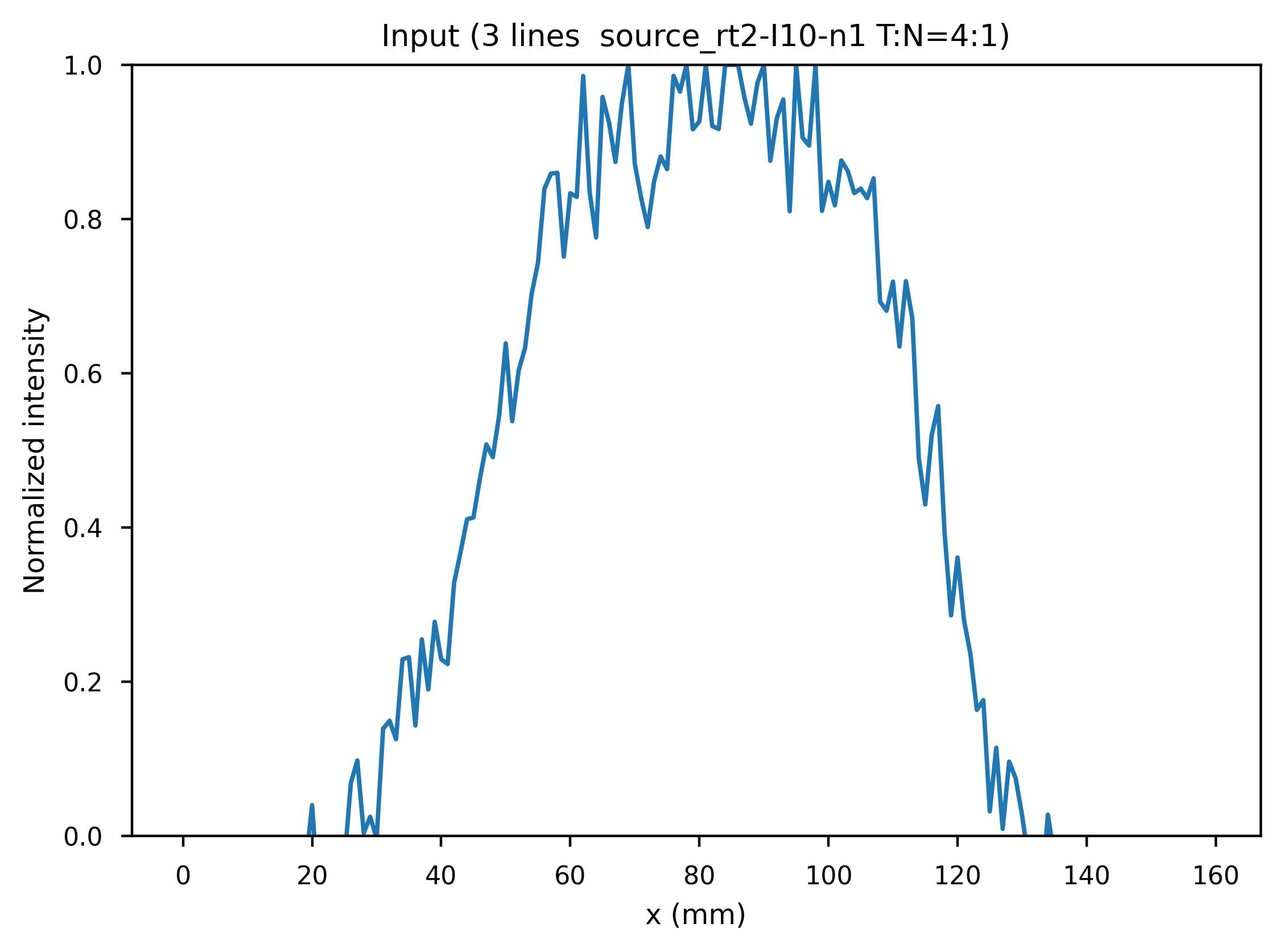}\label{input_3}}
	\caption{An example of network input. (a) XY gamma generation heatmap, (b) XZ gamma generation heatmap (c) YZ gamma generation heatmap and (d) normalized intensity as a function of $x$.}
	\label{fig:network_inputs}
\end{figure}

\begin{figure}[htbp]
	\centering
	\subfloat[][]{\includegraphics[width=0.4\linewidth]{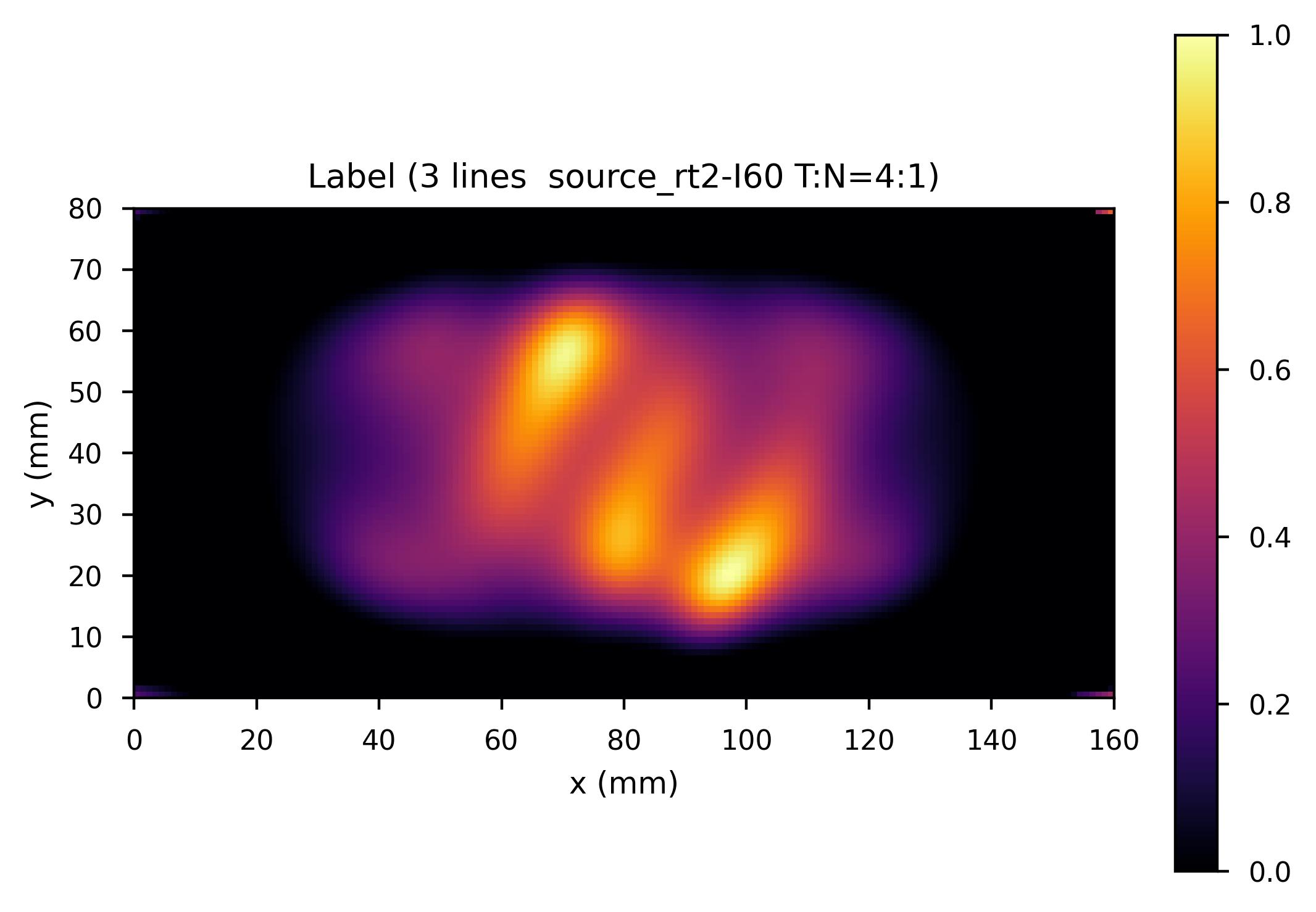}\label{label_0}} \quad
    \subfloat[][]{\includegraphics[width=0.4\linewidth]{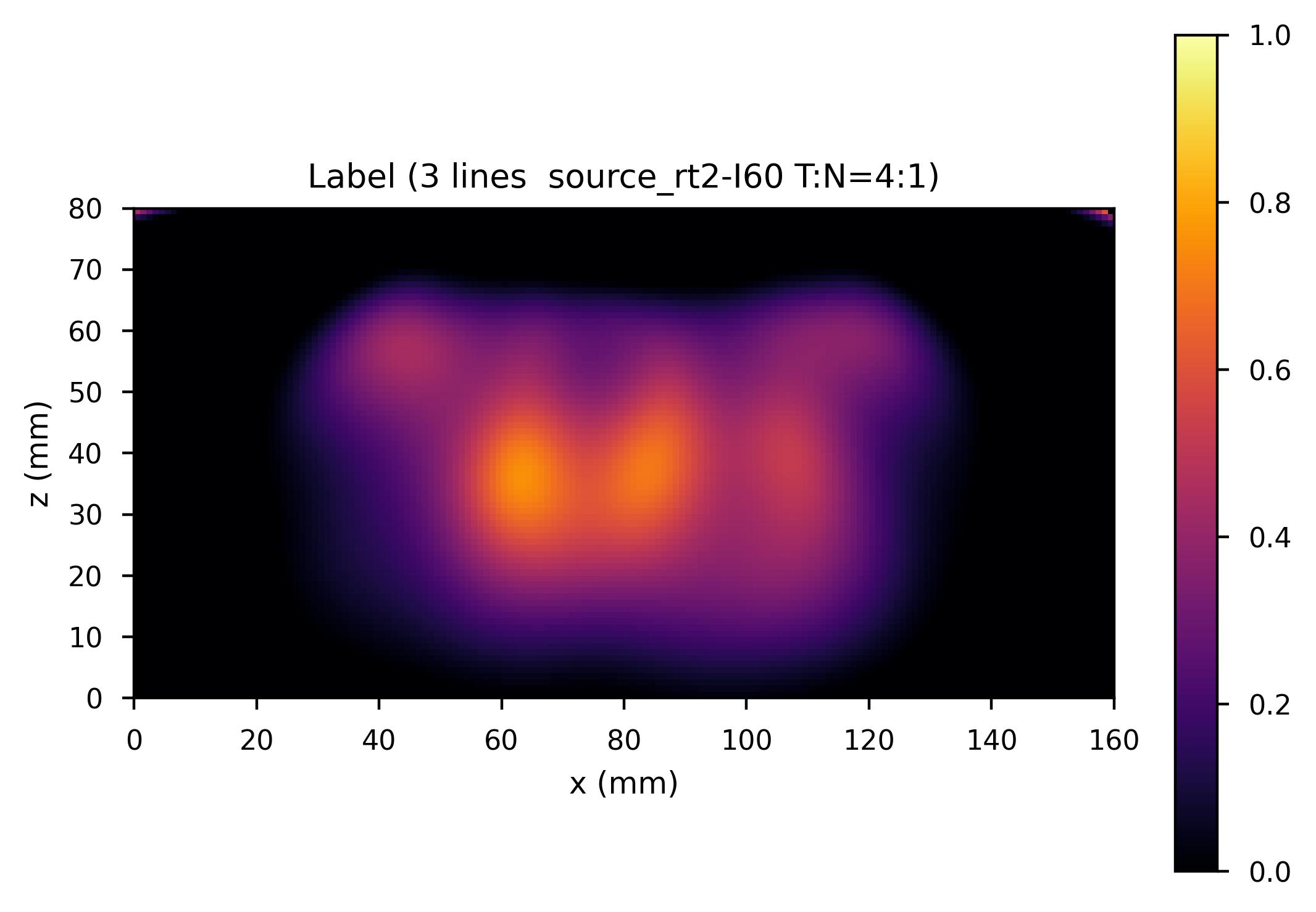}\label{label_1}} \quad
    \subfloat[][]{\includegraphics[width=0.4\linewidth]{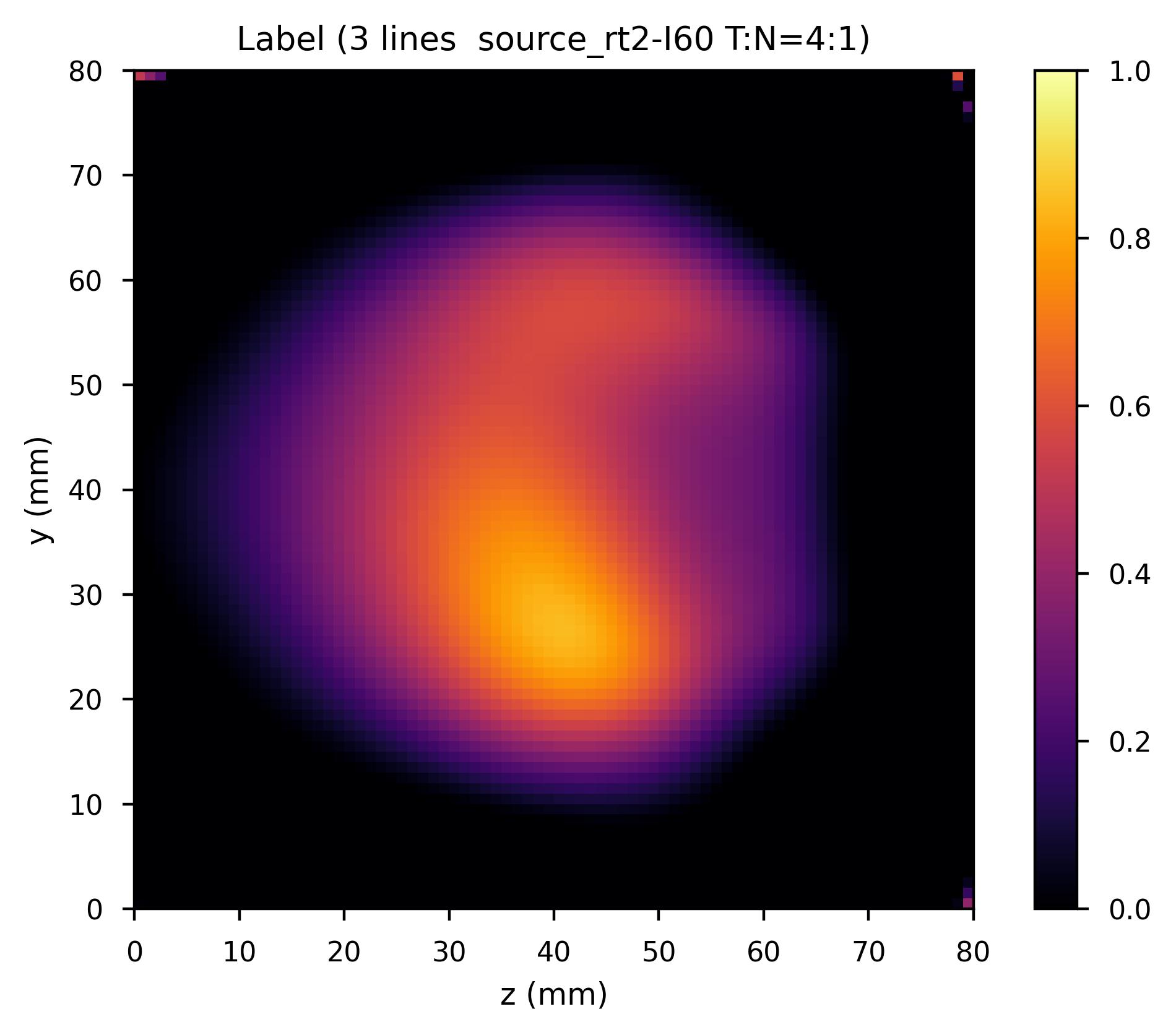}\label{label_2}} \quad
	\subfloat[][]{\includegraphics[width=0.4\linewidth]{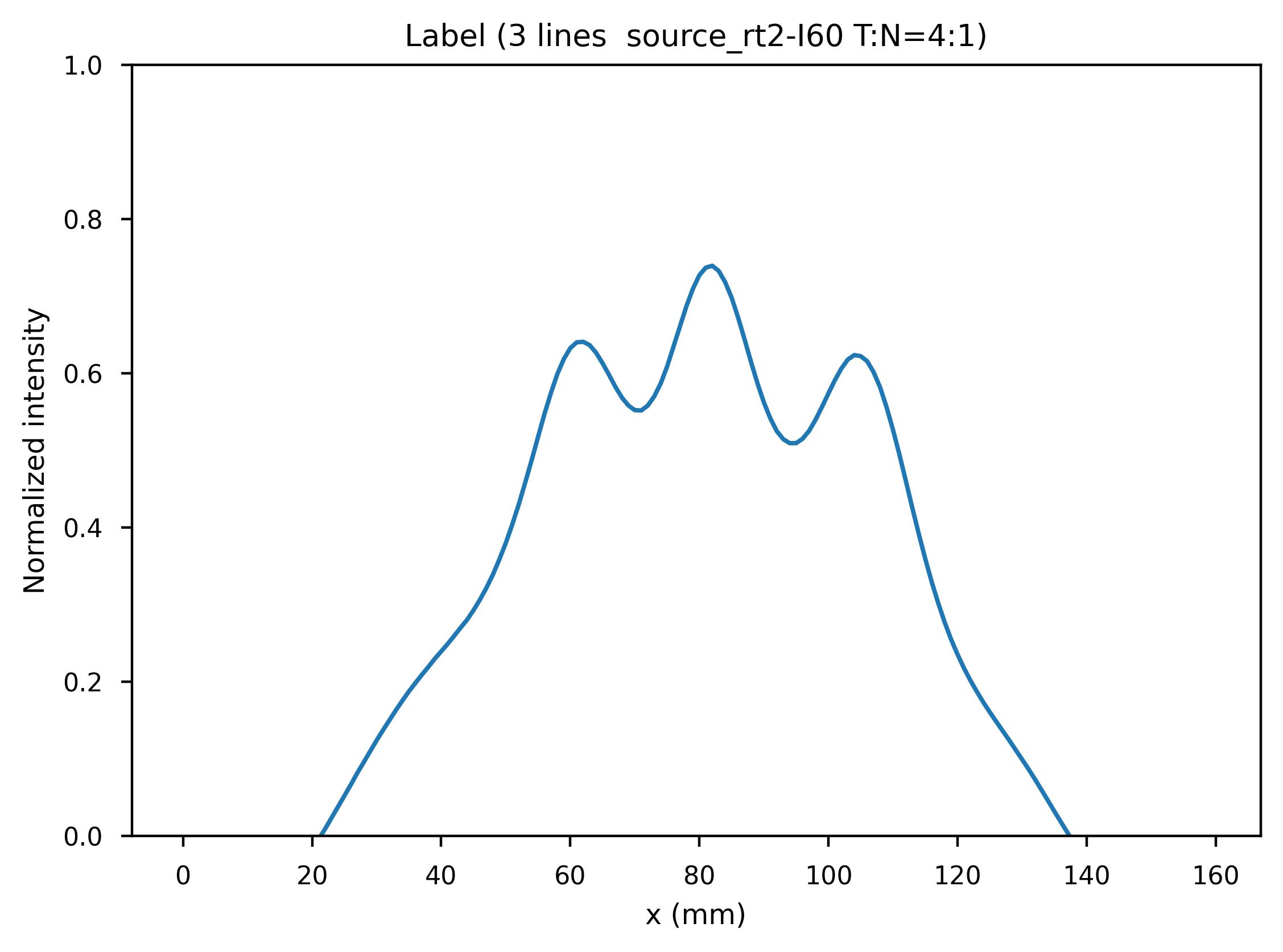}\label{label_3}}
	\caption{Corresponding label. (a) XY gamma generation heatmap, (b) XZ gamma generation heatmap (c) YZ gamma generation heatmap and (d) normalized intensity as a function of $x$.}
	\label{fig:network_labels}
\end{figure}

The dataset consists in the output of the $10$th iteration of the list-mode MLEM algorithm (input images) and the output of the $60$th iteration (label images) of the $71$ simulated gamma source distributions and of the data augmentation gamma source distributions obtained by applying four different rototranslations, for a total of $355$ samples with the corresponding labels. In order to further increase the size of the dataset, $41$ noisy images were created by adding one out of four different levels of Gaussian white noise to each of the sample images (if $\Delta{}\tilde{f}$ denotes the difference between the maximum and minimum value of the input image the standard deviations of the four levels are: $\sigma_{0}=\Delta{}\tilde{f}/16$, $\sigma_{1}=\Delta{}\tilde{f}/20$, $\sigma_{2}=\Delta{}\tilde{f}/30$, $\sigma_{3}=\Delta{}\tilde{f}/40$), leaving the corresponding label unchanged, resulting into a global dataset of $355\times{}42=14910$ input images with corresponding label. These were distributed with a proportion of $70:10:20$ among the training set ($11130$ images), validation set ($1260$ images) and test set ($2520$ images), mantaining class balance among sets and distributing different rototranslations into different sets, so that every set contains almost new gamma source distributions with respect to the others. Fig~\ref{fig:network_inputs} shows an example of input image, with noise level $1$ and rototranslation $2$. The corresponding label is displayed in Fig~\ref{fig:network_labels}.

\par{}
The 3D standard U-Net, 3D dual frame U-Net and 3D tight frame U-Net are described in sections~\ref{subsec:unets} and represented in Fig.~\ref{fig:architectures_2D}(a)-(b) and Fig.~\ref{fig:tight_frame_3D} (they are 4D representations, the plane perpendicular to the page corresponds to three dimensional space). All the U-Nets include convolutional layers with $3\times{}3$ filters and rectified linear units (ReLU). The first two network employ average pooling and unpooling layers. The tight frame U-Net uses Haar wavelet decomposition with eight filters, the corresponding unpooling operation with eight synthesis filters \citep{mallat,damelin_mathematics_signal} and the weighted addition of nine input tensors with learnable weights. All networks include skip connections. Every input image and label is normalized to the interval $[0,1]$.
\par{}
The networks were trained using ADAM algorithm \citep{sayed_22} with a learning rate equal to $0.001$. The loss function was the normalize mean square error (NMSE), defined below. The batch size was set equal to $1$ owing to the large size of three-dimensional images. Data was lazy-loaded to the main memory and then asynchronously loaded to the GPU.
The networks were implemented using PyTorch\footnote{An optimized tensor library for deep learning using GPUs and CPUs, based on an automatic differentiation system \citep{automatic_differentiation}.}. A A100 PCIe $40$ $GB$ GPU with Ampere architecture and eight AMD EPYC 7742 64-Core CPUs ($2.25$ $GHz$) were used. All three networks require about $7-8$ days for training. The number of training epochs $n_{te}$ (with epochs running from $0$ to $n_{te}-1$) and best epoch with its training and validation average NMSEs are reported in Table~\ref{table_best_epochs}. The best epoch model was used for validation.
\begin{table}[htbp]
    \centering
    \begin{tabular}{|p{3.2cm}||p{0.3cm}|p{1.8cm}|p{1.6cm}|p{1.8cm}|} 
    \hline
    Network & $n_{te}$ & Best epoch & Tr. NMSE & Val. NMSE \\ [0.5ex] 
    \hline
    U-Net & 56 & 39 & 0.03396 & 0.02865 \\ 
    Dual frame U-Net & 53 & 50 & 0.03280 & 0.02571 \\
    Tight frame U-Net & 52 & 48 & 0.01102 & 0.01113\\ [1ex] 
    \hline
    \end{tabular}
    \caption{Number of training epochs $n_{te}$ and best epoch with its training and validation NMSEs for each network.}
    \label{table_best_epochs}
\end{table}
\par{}
For quantitative evaluation, three different metrics were used: the \emph{normalized mean square error} (NMSE) value defined as
\begin{eqnarray}
	NMSE = \frac{\| f^* - \hat{f}\|_{2}^{2}}{\| f^*\|_{2}^{2}} = \frac{\sum_{i=1}^{M} \sum_{j=1}^{N} \sum_{k=1}^{O} [f^*(i,j,k) - \hat{f}(i, j, k)]^2}{\sum_{i=1}^{M}\sum_{j=1}^{N}\sum_{k=1}^{O}[f^*(i,j,k)]^2},
\end{eqnarray}
where $M$, $N$ and $O$ are the number of pixels in the $x-$, $y-$ and $z-$direction and $\hat{f}$ and $f^*$ denote the reconstructed images and labels, respectively; the \emph{peak signal to noise ratio} (PSNR), defined by\footnote{The \emph{mean square error} (MSE) is defined as $MSE = \| f^* - \hat{f}\|_{2}^{2}/(NMO)$. $\|f^*\|_\infty{}=|max_{(i,j,k)}f^*(i,j,k)|$.}
\begin{eqnarray}
	PSNR= 10 \log_{10}\left( \frac{\|f^*\|_\infty{}^{2}}{MSE}\right) = 20 \cdot \log_{10} \left(\frac{\sqrt{NMO}\|f^*\|_\infty}{\|\hat f- f^*\|_2}\right) \  ;
\label{eq:psnr}		 
\end{eqnarray}
The \emph{structural similarity} index measure (SSIM) \citep{wang_image_quality}, defined as
\begin{equation}
	SSIM = \dfrac{(2\mu_{\hat f}\mu_{f^*}+c_1)(2\sigma_{\hat f f^*}+c_2)}{(\mu_{\hat f}^2+\mu_{f^*}^2+c_1)(\sigma_{\hat f}^2+\sigma_{f^*}^2+c_2)},
\end{equation}
where $\mu_{\hat f}$ is a average of $\hat f$, $\sigma_{\hat f}^2$ is a variance of $\hat f$ and $\sigma_{\hat f f^*}$ is a covariance of $\hat f$ and $f^*$. While the first two metrics quantify the difference in the values of the corresponding pixels of the reference and reconstructed images, the structural similarity index quantifies the similarity based on luminance, contrast and structural information, similarly to the human visual perception system. Low values of $NMSE$ and high values of $PSNR$ indicate similarity between images. The structural similarity index takes values in $[-1,1]$, where values close to one indicate similarity, zero indicates no similarity and values close to $-1$ indicate anti-correlation.

\section{Results}
\label{subsec:unets_results}

\begin{figure}[htbp]
	\centering
	\subfloat[][]{\includegraphics[width=0.4\linewidth]{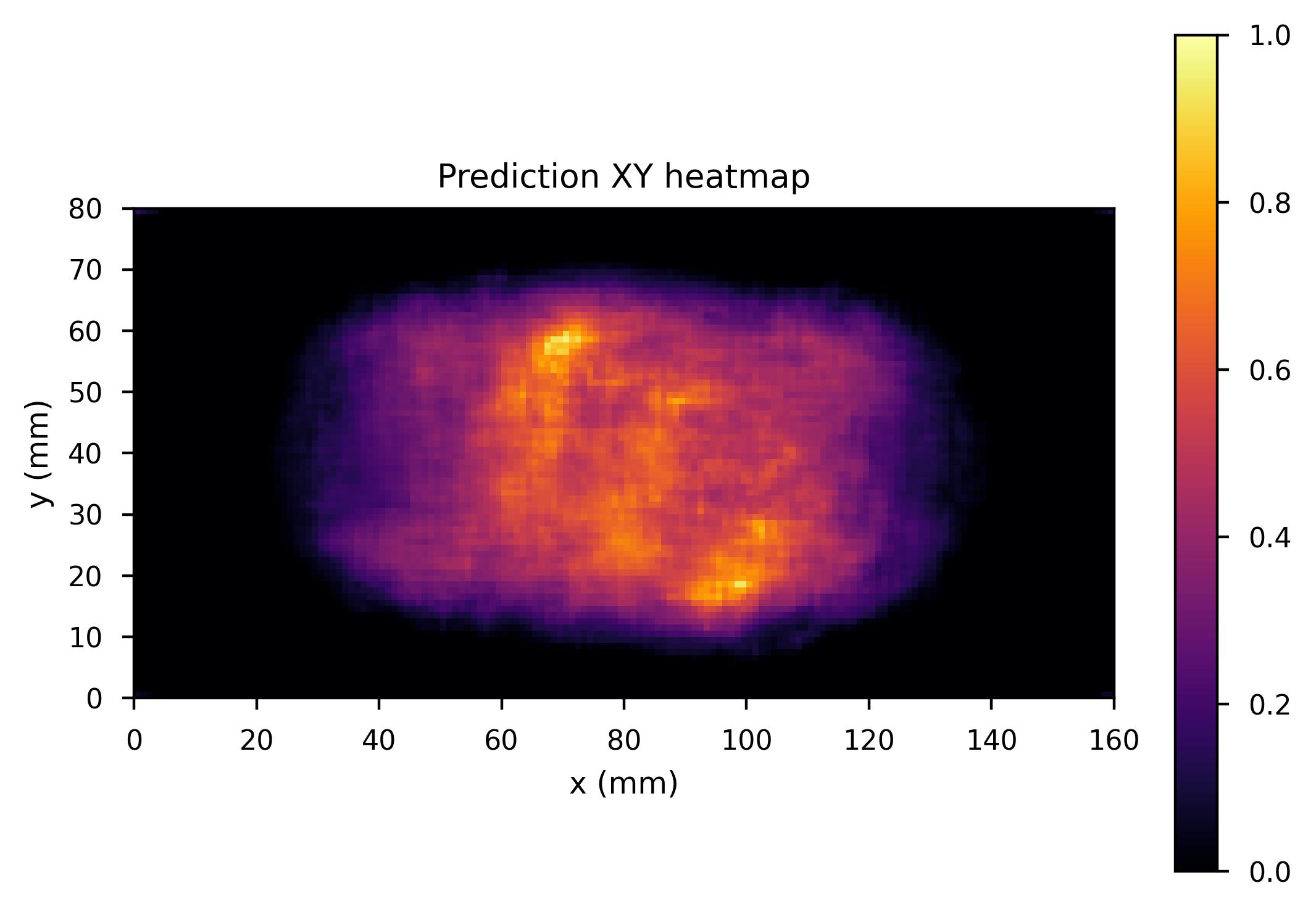}\label{unet_pred_0}} \quad
    \subfloat[][]{\includegraphics[width=0.4\linewidth]{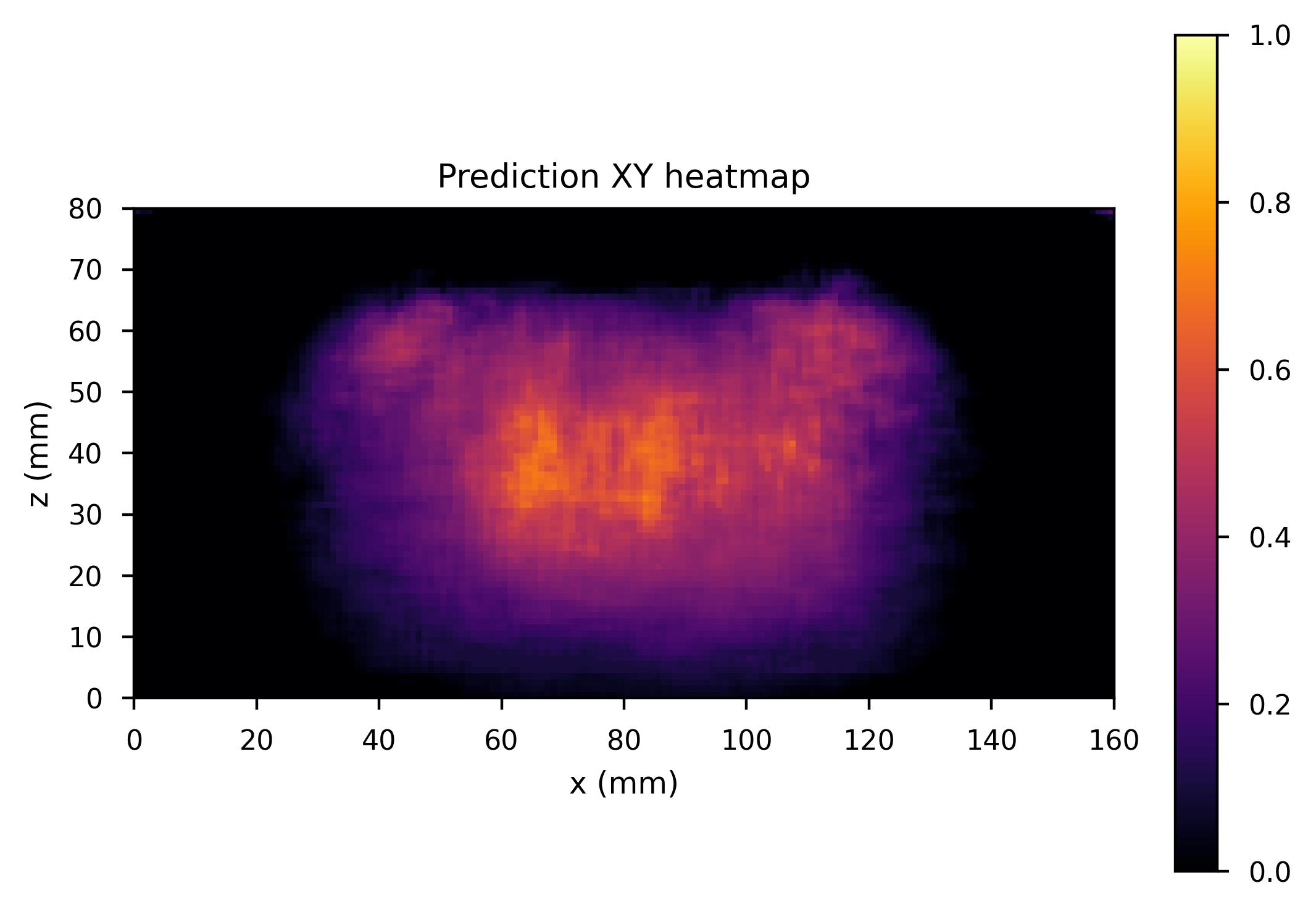}\label{unet_pred_1}} \quad
    \subfloat[][]{\includegraphics[width=0.4\linewidth]{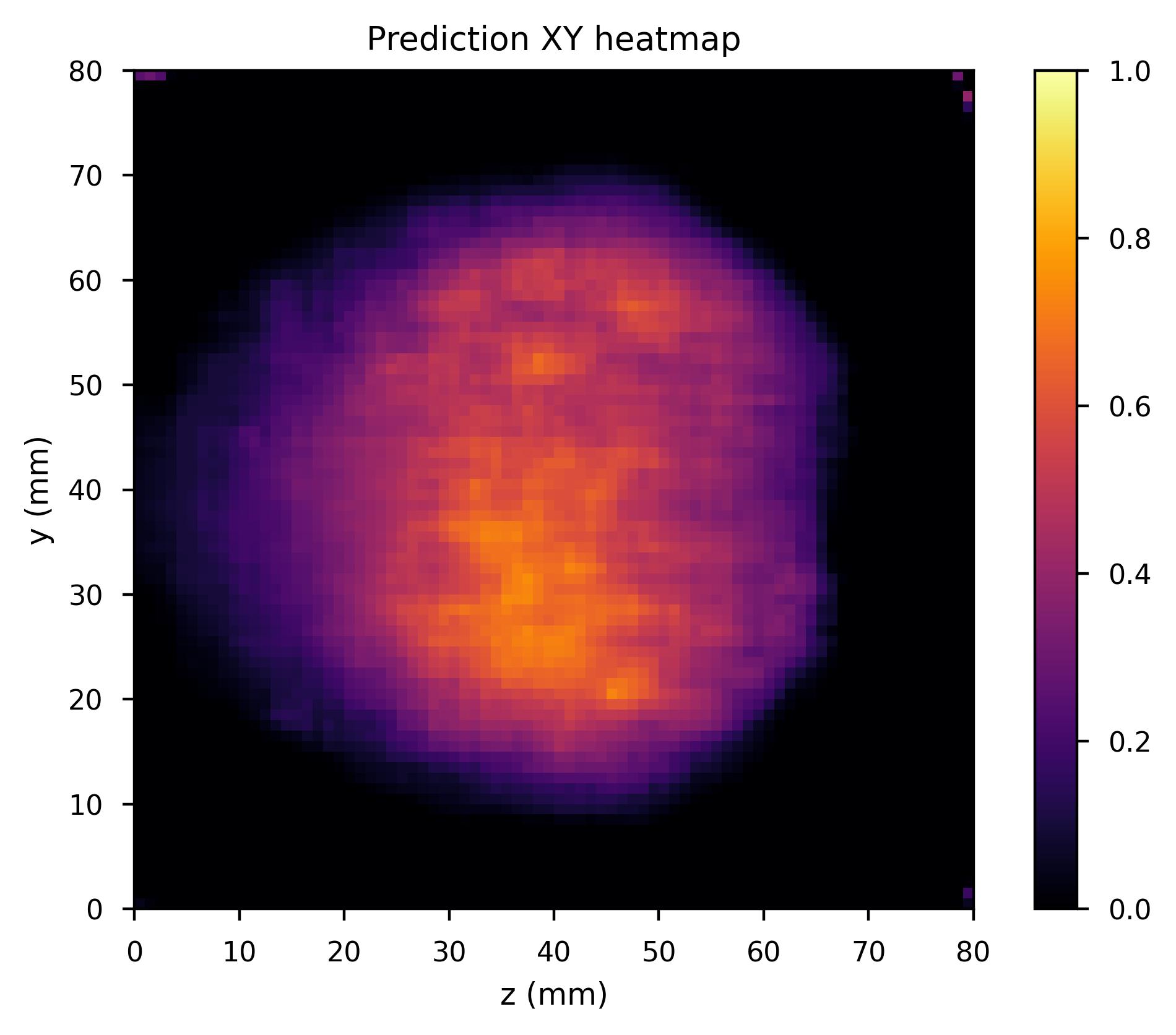}\label{unet_pred_2}} \quad
	\subfloat[][]{\includegraphics[width=.4\linewidth]{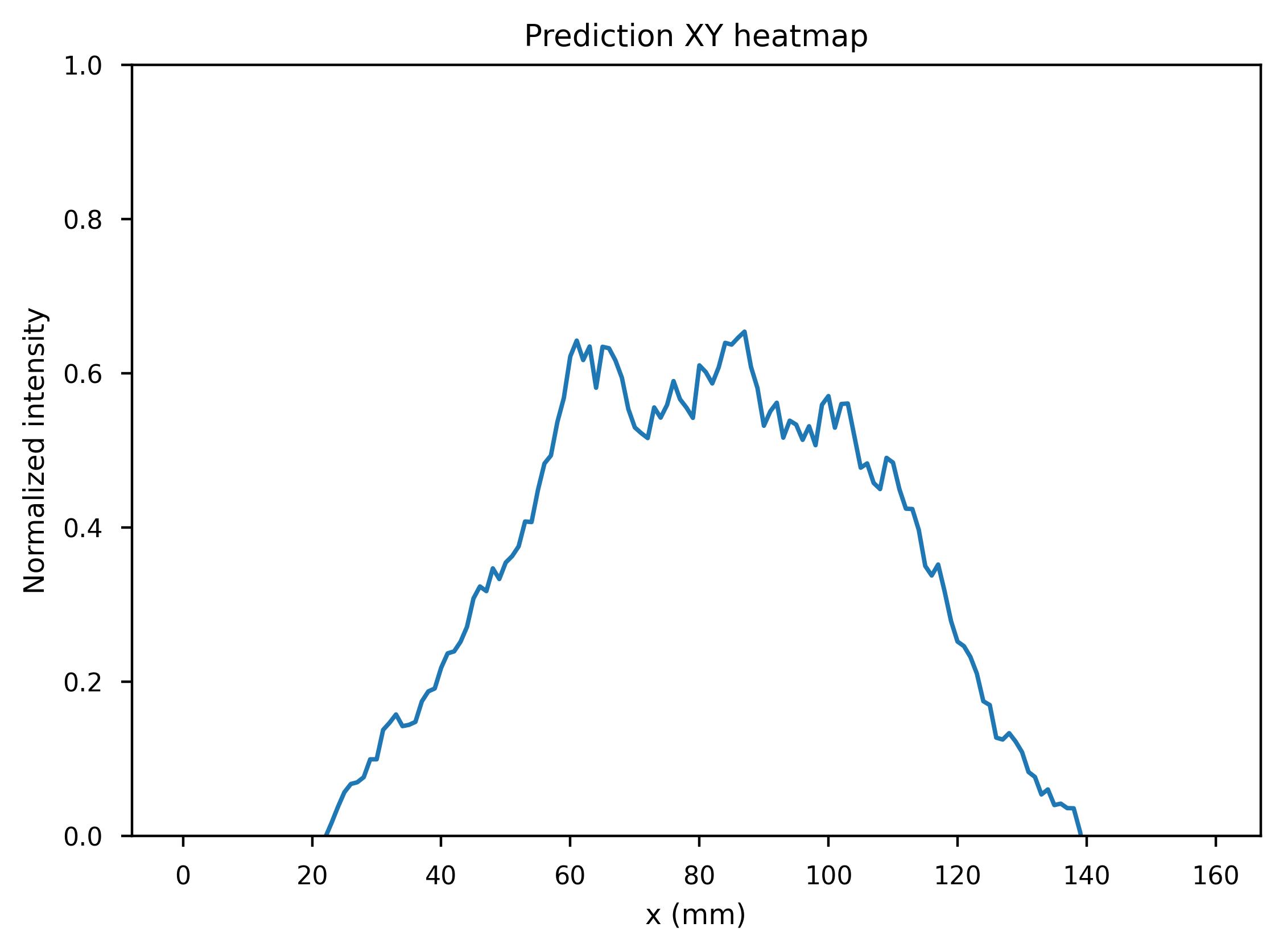}\label{unet_pred_3}}
	\caption{U-Net predictions (a) XY gamma generation heatmap, (b) XZ gamma generation heatmap (c) YZ gamma generation heatmap and (d) normalized intensity as a function of $x$.}
	\label{fig:unet_predictions}
\end{figure}

\begin{figure}[htbp]
	\centering
	\subfloat[][]{\includegraphics[width=0.4\linewidth]{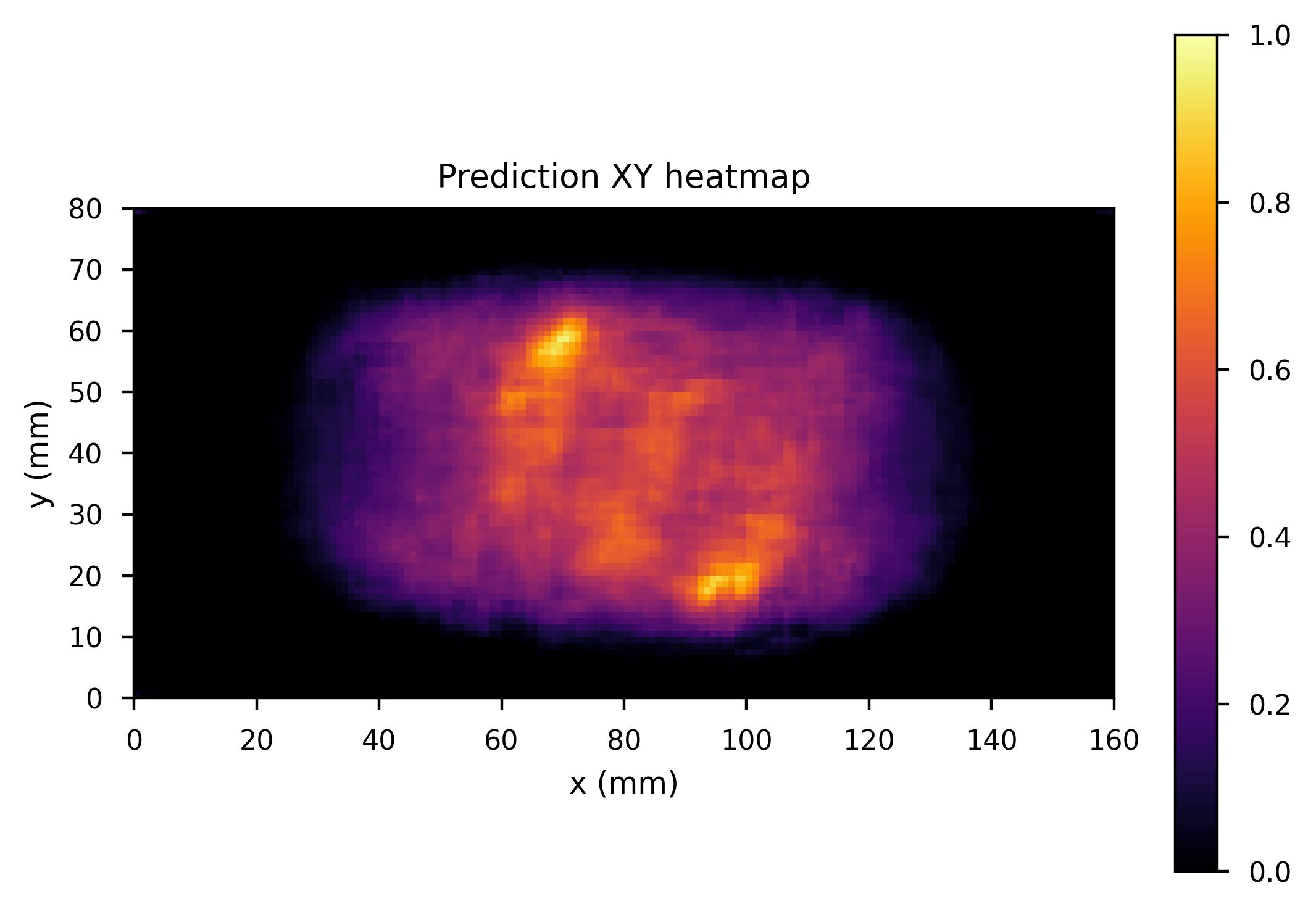}\label{dual_pred_0}} \quad
    \subfloat[][]{\includegraphics[width=0.4\linewidth]{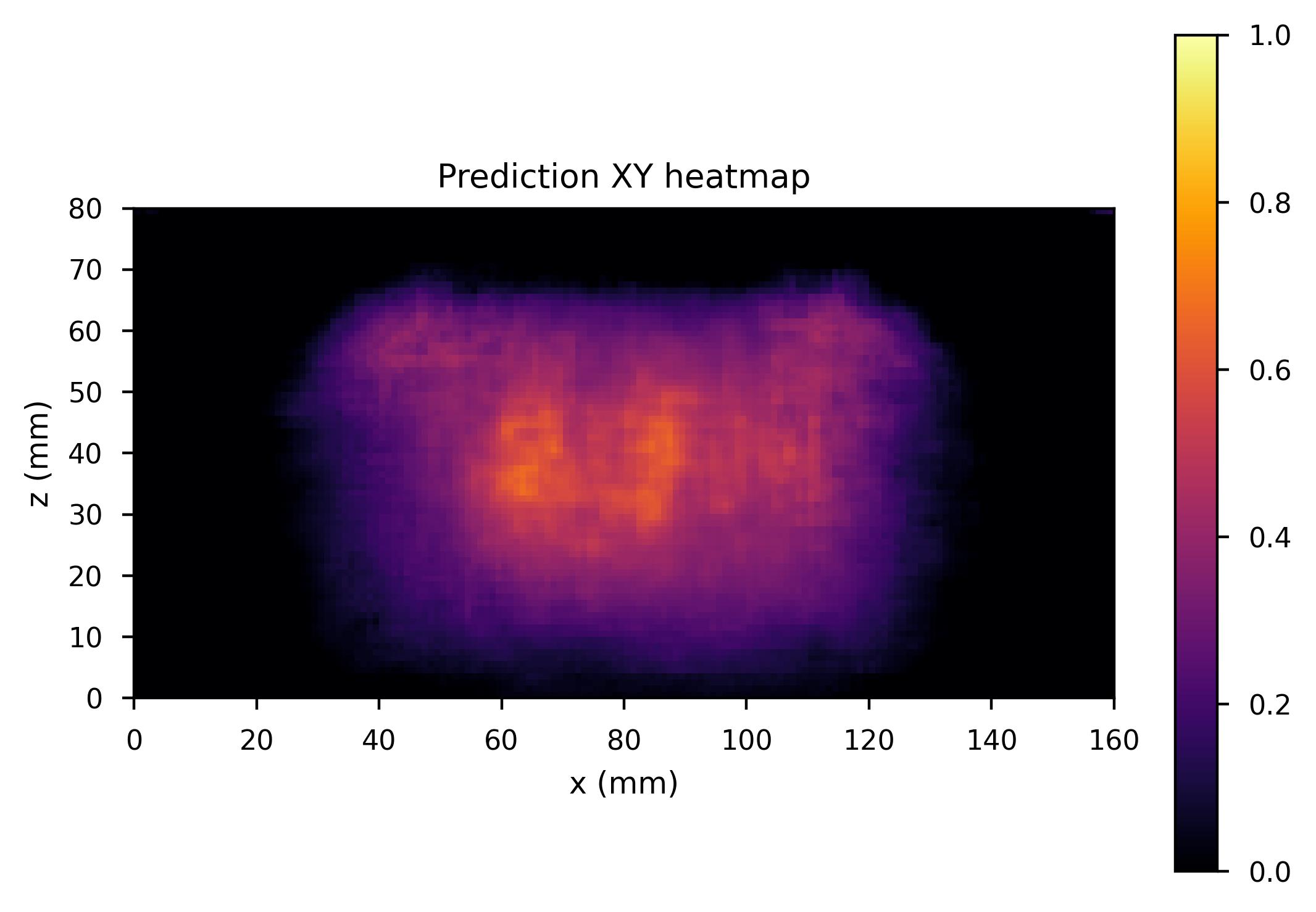}\label{dual_pred_1}} \quad
    \subfloat[][]{\includegraphics[width=0.4\linewidth]{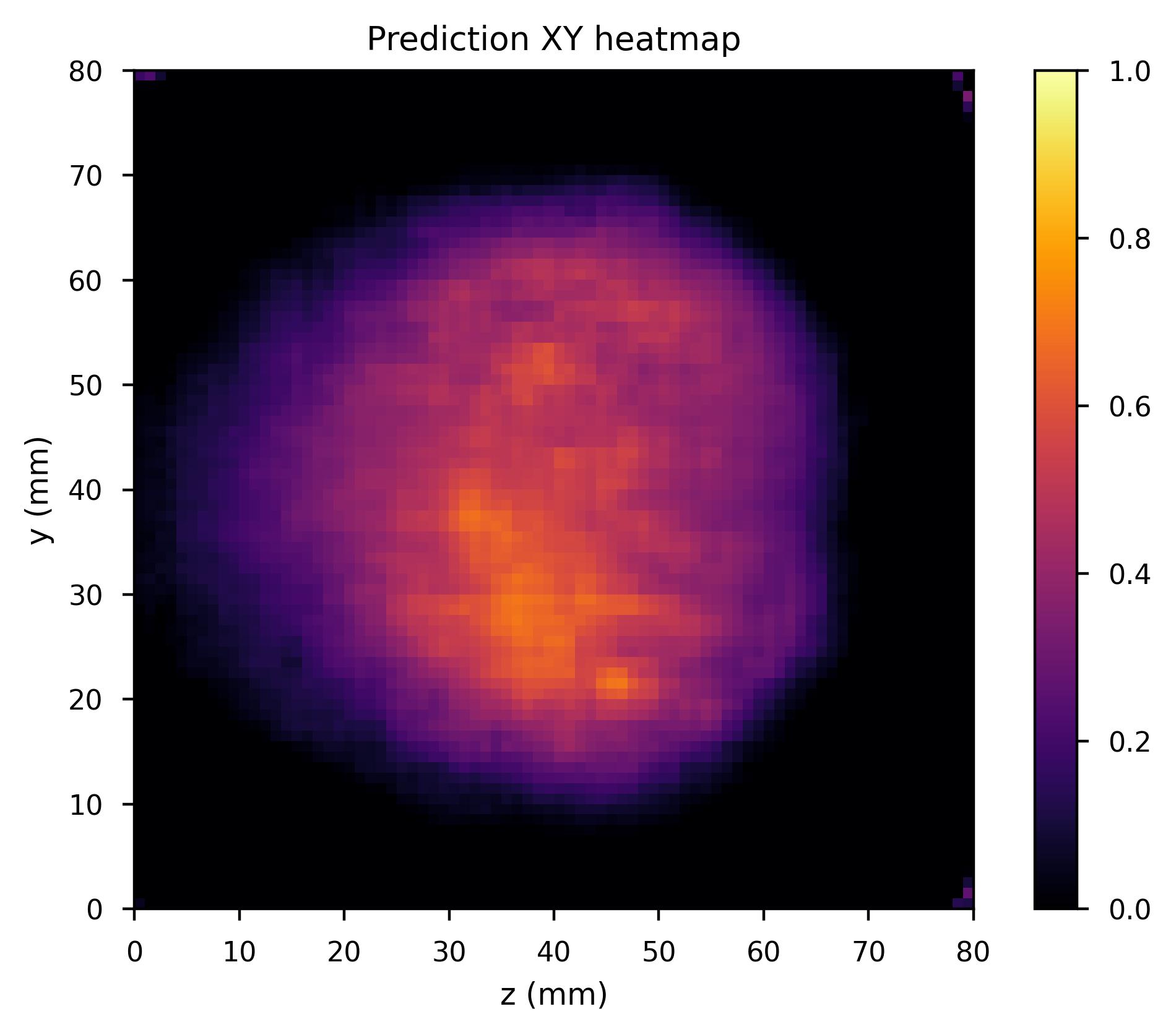}\label{dual_pred_2}} \quad
	\subfloat[][]{\includegraphics[width=0.4\linewidth]{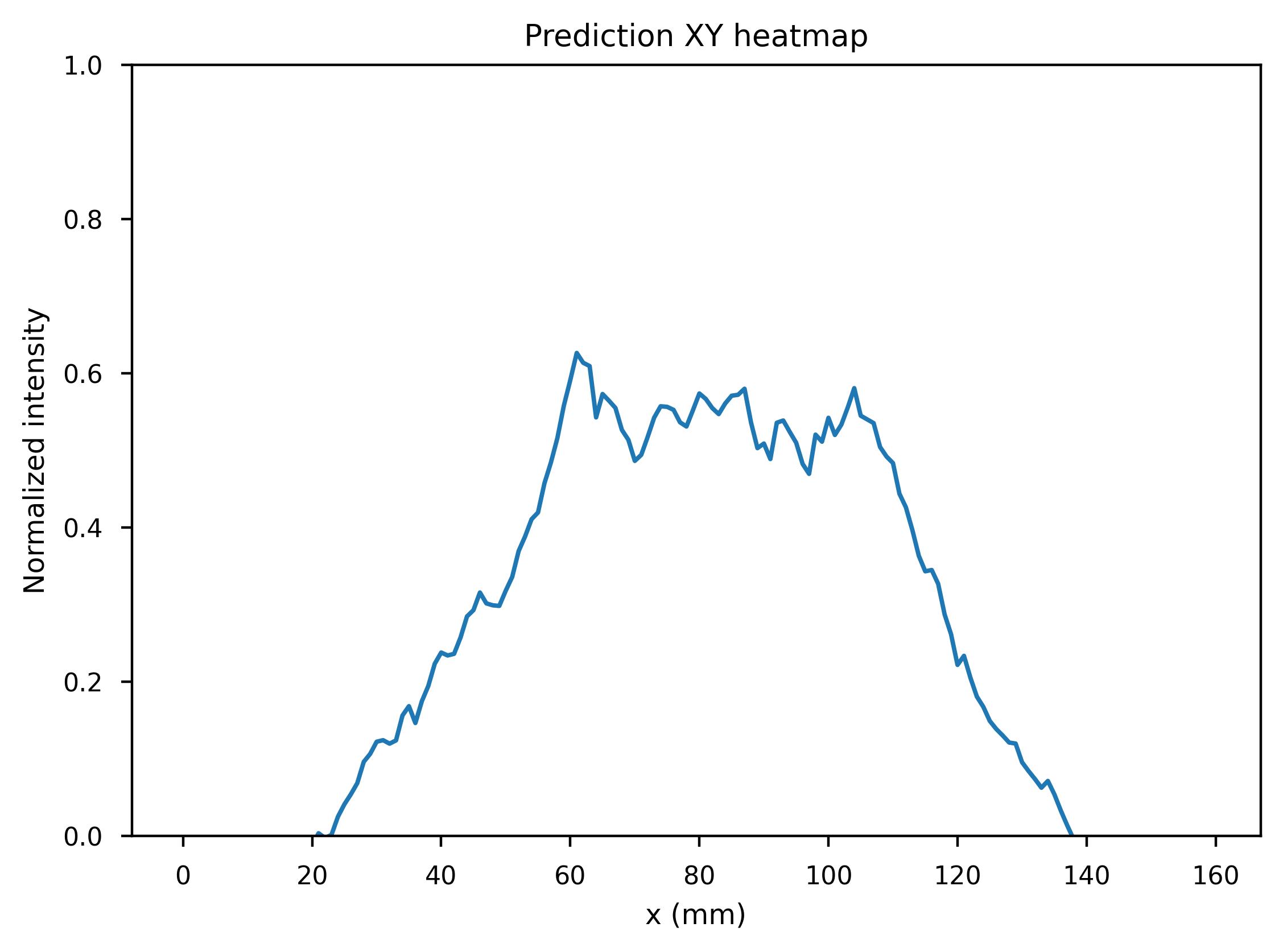}\label{dual_pred_3}}
	\caption{Dual frame U-Net predictions (a) XY gamma generation heatmap, (b) XZ gamma generation heatmap (c) YZ gamma generation heatmap and (d) normalized intensity as a function of $x$.}
	\label{fig:dual_predictions}
\end{figure}

\begin{figure}[htbp]
	\centering
	\subfloat[][]{\includegraphics[width=0.4\linewidth]{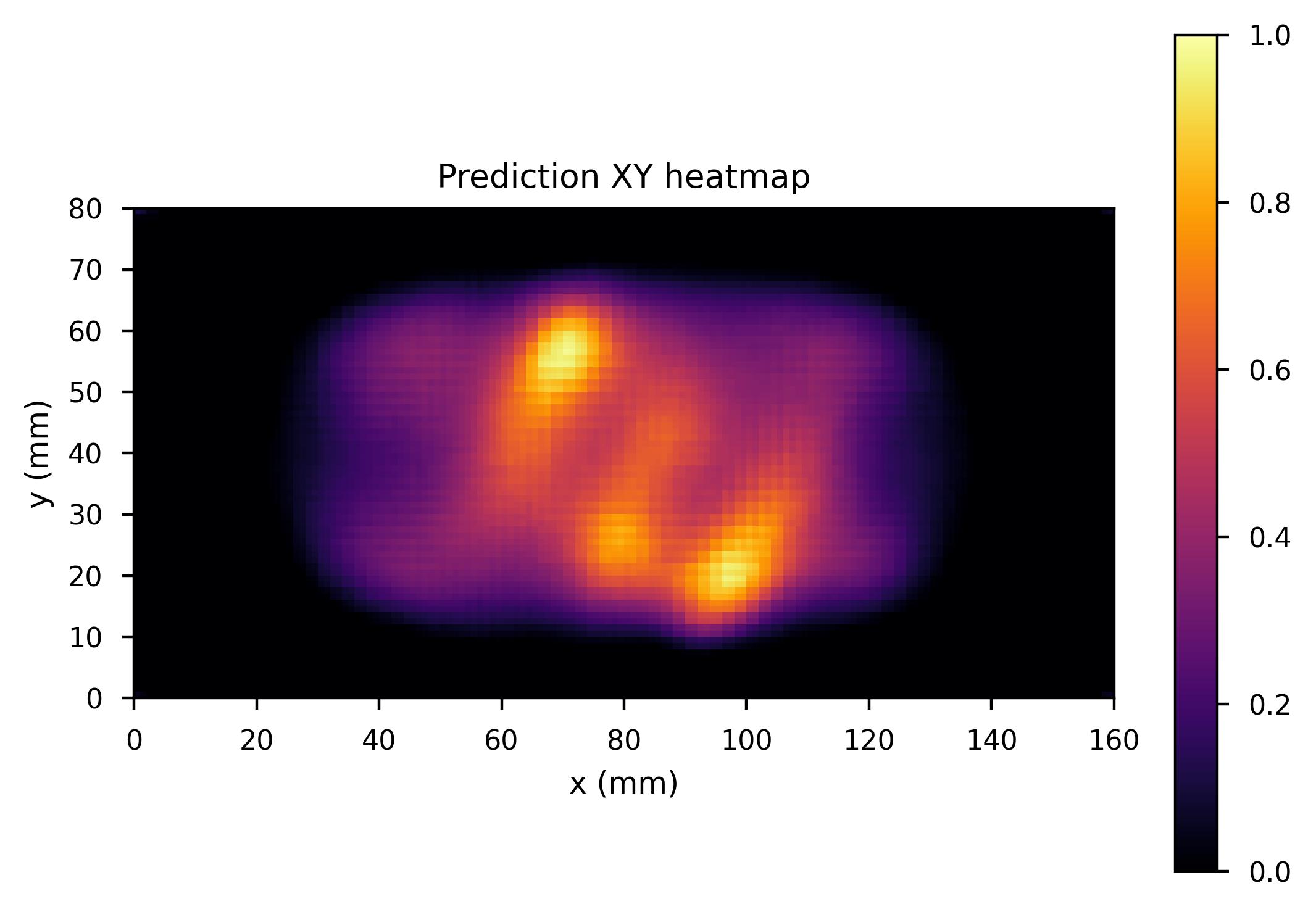}\label{tight_pred_0}} \quad
    \subfloat[][]{\includegraphics[width=0.4\linewidth]{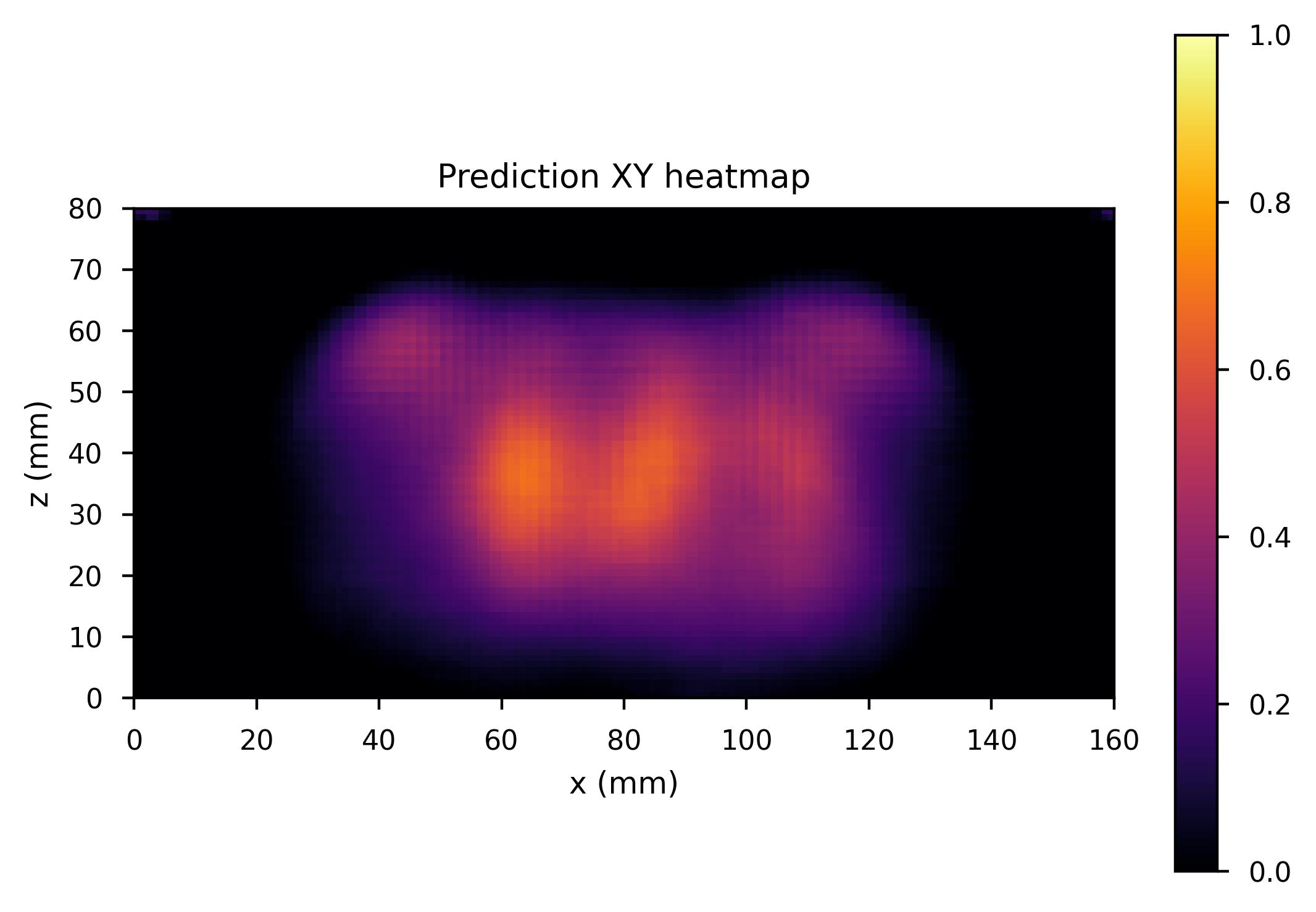}\label{tight_pred_1}} \quad
    \subfloat[][]{\includegraphics[width=0.4\linewidth]{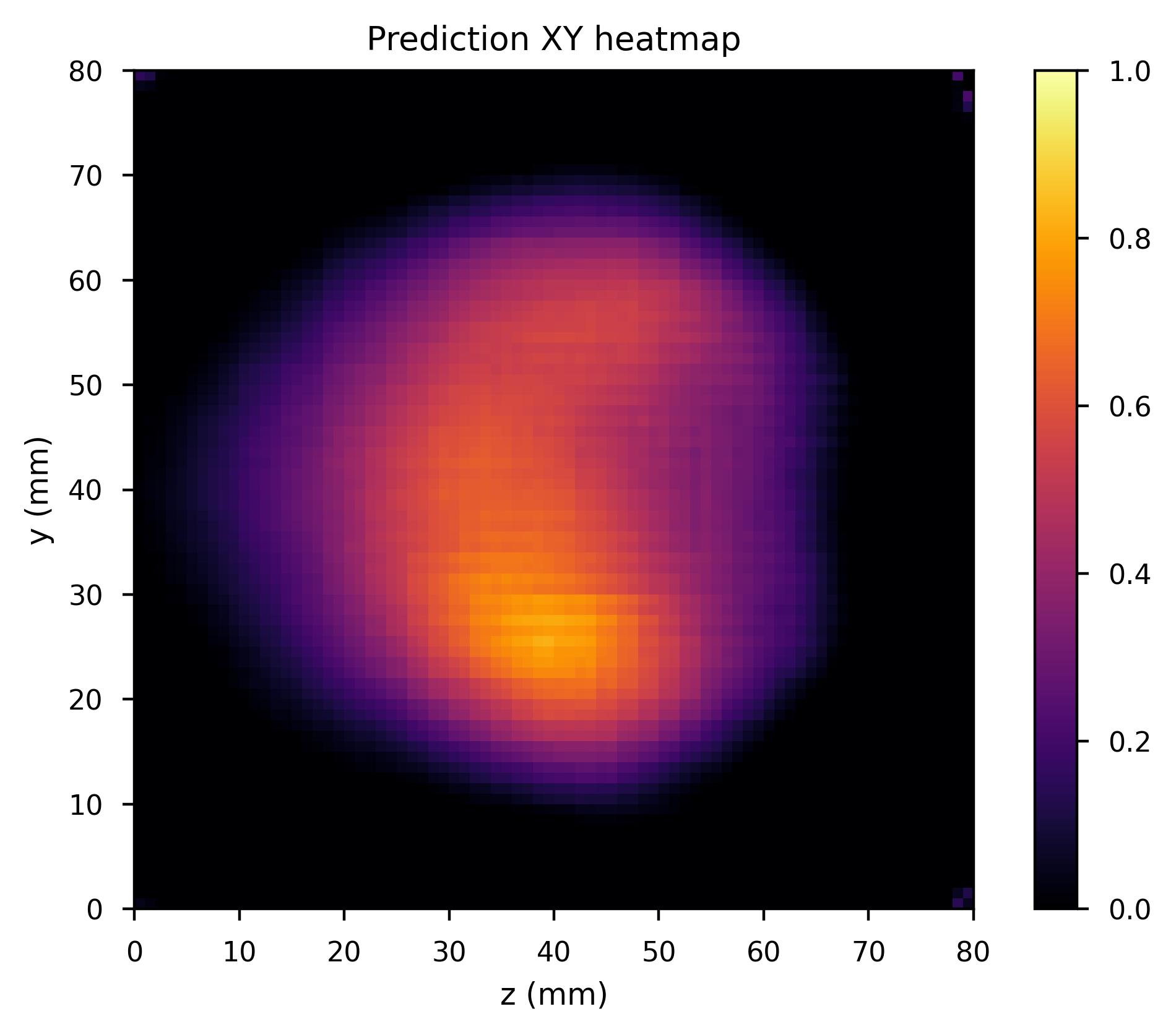}\label{tight_pred_2}} \quad
	\subfloat[][]{\includegraphics[width=0.4\linewidth]{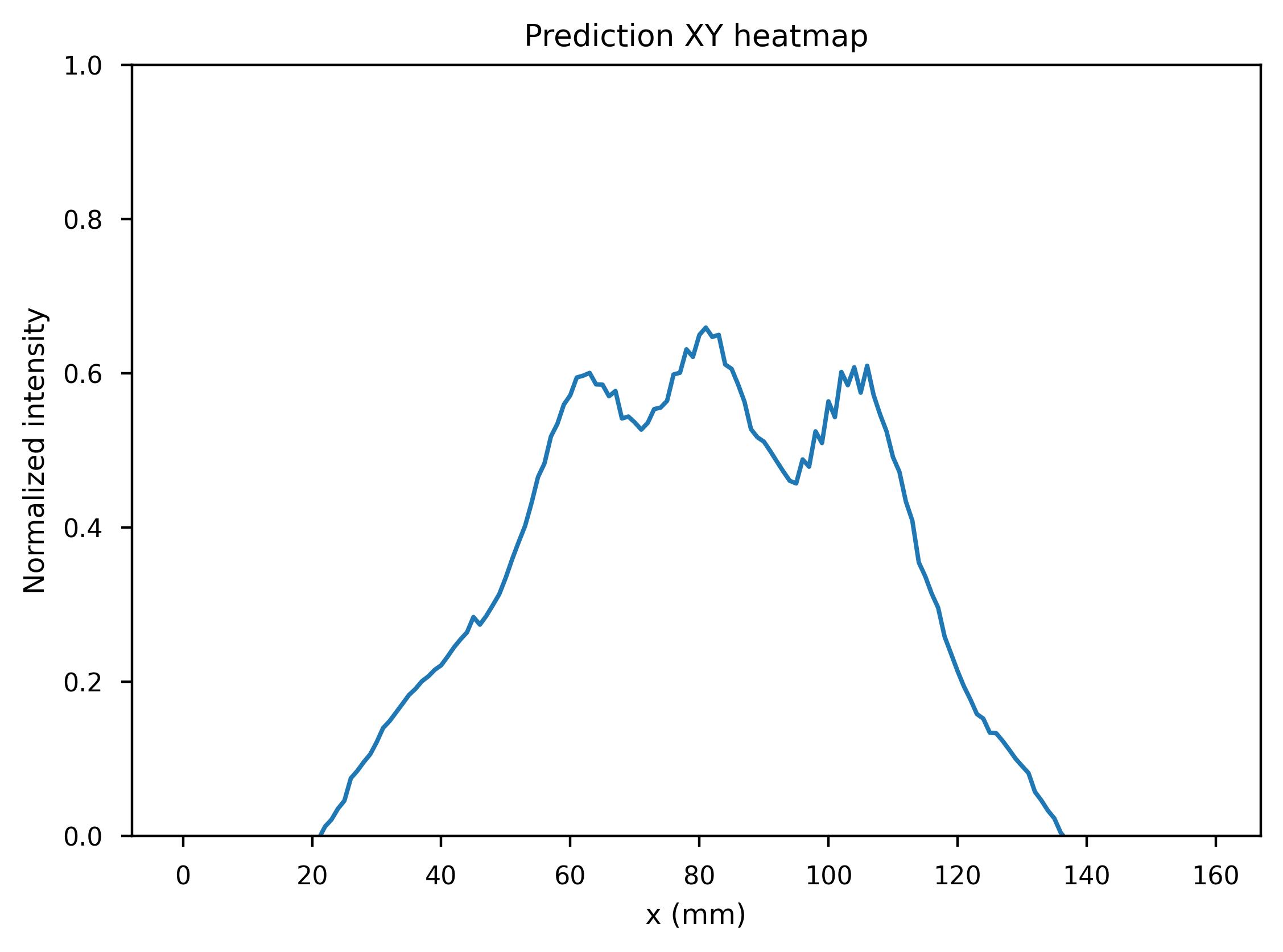}\label{tight_pred_3}}
	\caption{Tight frame U-Net predictions (a) XY gamma generation heatmap, (b) XZ gamma generation heatmap (c) YZ gamma generation heatmap and (d) normalized intensity as a function of $x$.}
	\label{fig:tight_predictions}
\end{figure}

Fig~\ref{fig:unet_predictions}, Fig~\ref{fig:dual_predictions} and Fig~\ref{fig:tight_predictions} show the predicted reconstructions of the three models given the input in Fig~\ref{fig:network_inputs}. It can be observed that while the standard U-Net and the dual frame U-Net are more affected by degradations, the tight frame U-Net produces a prediction very similar to the label image in Fig~\ref{fig:network_labels} even if only two levels are considered in the architecture (Fig~\ref{fig:tight_frame_3D}). This observations are confirmed by similarity metrics reported in Table~\ref{table_unets_results}.
\begin{table}[htbp]   
    \centering
    \begin{tabular}{|p{3.2cm}||p{1.5cm}|p{1.5cm}|p{1.5cm}|} 
    \hline
     & NMSE & PSNR & SSIM \\ [0.5ex] 
    \hline
    Standard U-Net & $0.031803$ & $36.119379$ & $0.754417$ \\ 
    Dual frame U-Net & $0.029113$  & $36.396008$ & $0.726286$  \\
    Tight frame U-Net & $0.011953$  & $40.615813$ & $0.853548$\\ [1ex] 
    \hline
    \end{tabular}
    \caption{Average NMSE, PSNR and SSIM on the test set for the trained U-Net, dual frame U-Net and tight frame U-Net modules.}
    \label{table_unets_results} 
\end{table}
The performance of the standard U-Net and the dual frame U-Net are essentially comparable, with the latter performing slightly better in terms of NMSE and PSNR but slightly worse in terms of SSIM. The performance of the two networks and the presence of degradations can be explained by considering that the standard U-Net doesn't satisfy the frame condition and the dual frame U-Net, while satisfying the frame condition, tends to amplify noise. The tight frame architecture showed a considerable improvement in performance considering all three metrics, and in particular in the SSIM, which quantifies structural information.
\par{}
In terms of processing time, U-Nets regression takes generally less than a second, so that the overall reconstruction time is dominated by the time necessary to obtain the input image, which is of the order of $4-6$ minutes. Considering a BNCT treatment duration of $30-90$ minutes, the obtained reconstruction time performance represents a significant improvement compared to classical iterative methods (processing time is reduced by a factor of about $6$ with respect to the $24-36$ minutes needed for $60$ iterations in the case of list-mode MLEM), making this kind of approach a valid step towards real time dose monitoring during BNCT treatment.
\par{}
Notice that although the source geometries created in the simulation are not biologically realistic, the same procedure can be applied with distributions obtained in real medical practice. Moreover transfer learning techniques \citep{murphy_22} could be employed to take advantage of the training already done with simulated datasets.

\section{Conclusions}

Boron Neutron Capture Therapy (BNCT) represents a promising form of cancer therapy because of its high selectivity towards cancer tissue. However at present there are no viable imaging methods capable of in vivo monitoring dose during treatment. Compared to other imaging techniques under investigation, Compton imaging offers various advantages, but the main difficulty in this type of approach is the complexity of Compton image reconstruction, which is associated with long reconstruction times, comparable with BNCT treatment duration. This calls for the development of new reconstruction techniques with lower computational cost.
\par{}
In order to investigate the potentialities of Compton imaging with CZT detectors for BNCT, a Geant4 simulation of a simplified detector in a BNCT setting has been implemented, considering several tumor region geometries in order to produce a large enough dataset for the training phase of deep neural network algorithms.  
\par{}
In order to reduce reconstruction time, the U-Net architecture and two variants based on the deep convolutional framelets framework, the dual frame U-Net and the tight frame U-Net, were applied to reduce degradation in few-iterations reconstructed images. Encouraging results were obtained both in terms of visual inspection and in terms of the three metrics used to evaluate the similarity with the reference images (NMSE, PSNR and SSIM), especially with the use of tight frame U-Nets. The lower performance of standard U-Net architecture and of the dual frame variant was attributed to the fact that the former doesn't satisfy frame condition, while the latter tends to amplify noise. The processing time was reduced on average by a factor of about $6$ with respect to classical iterative algorithms, with most it amounting to the starting image reconstruction time of about $4-6$ minutes. This can be considered a good reconstruction time performance, considering typical BNCT treatment times.
\par{}
In principle it would be possible to further improve reconstruction accuracy and reduce processing time by improving quality and time performance in the reconstruction of the input image provided to the U-Net, for example by employing unrolled optimization algorithms.

\vspace{6pt}

\dataavailability{The data can be shared up on request.} 





\conflictsofinterest{The authors declare no conflicts of interest.} 

\begin{adjustwidth}{-\extralength}{0cm}

\reftitle{References}

\end{adjustwidth}
\end{document}